\documentclass{aa}  
\usepackage{graphicx}    
\usepackage{qtree221}

\usepackage{txfonts}
\usepackage{amssymb}
\usepackage{natbib}
\bibpunct{(}{)}{;}{a}{,}{,}
\usepackage{psfig}
\usepackage{slashbox}
\usepackage{times}
\usepackage{deluxetable}
\usepackage[figuresright]{rotating}

\newcommand{\lithiumline}{Li\,{\footnotesize I} 6707.8\,$\AA$}
\newcommand{\nallobserved}{332}
\newcommand{\nbinaries}{53}
\newcommand{\nxbinaries}{52}

\newcommand{\nsingles}{279}
\newcommand{\nrvsingles}{190}
\newcommand{\nxsingles}{187}

\newcommand{\nlisingles}{195}
\newcommand{\nlicert}{176}

\newcommand{\nsbuno}{8}

\newcommand{\nrvlinoxsingles}{10}

\newcommand{\nmembersingles}{184}
\newcommand{\npossmembersingles}{10}
\newcommand{\nrvmembdepleted}{3}
\newcommand{\nxmembdepleted}{7}
\newcommand{\nmembers}{237}

\newcommand{\nrvandxandlisingles}{158}
\newcommand{\nnoxsingles}{92}
\newcommand{\ndepleted}{8}
\newcommand{\ncontaminatings}{6}
\newcommand{\nbadspt}{7}
\newcommand{\uplimitvsini}{140} 
\newcommand{\clusterrvmean}{0.5}
\newcommand{\erclusterrvmean}{0.3}
\newcommand{\clusterstd}{3.2}
\newcommand{\erclusterstd}{0.2}
\newcommand{\fieldrvmean}{-45.2}
\newcommand{\erfieldrvmean}{10.5}
\newcommand{\fieldstd}{101.0}
\newcommand{\erfieldstd}{7.5}
\newcommand{\nhalphasample}{115}
\begin{document}
   \title{VLT/Flames observations of the star forming region NGC\,6530 }

   \subtitle{ }

   \author{L. Prisinzano \inst{1}
          \and
          F. Damiani\inst{1}
	  \and
	  G. Micela \inst{1}
	  \and
	  I. Pillitteri \inst{2}}

   \offprints{\email{loredana@astropa.inaf.it}}

   \institute{INAF - Osservatorio Astronomico di Palermo, Piazza del Parlamento 1, 90134
Palermo Italy \\
\and
Dipartimento di Scienze Fisiche ed Astronomiche, Universit\`a
di Palermo, Piazza del Parlamento 1, I-90134 Palermo Italy 
             }

   \date{Received 17/05/2006; accepted 03/10/2006}

 
  \abstract
    { Mechanisms regulating the evolution of pre-main sequence stars can be
   understood by studying stellar properties such as rotation, disk accretion,
   internal mixing and binarity. 
   To investigate such properties, we studied a
 sample of 332 candidate members of the massive and populous 
 star forming region
NGC\,6530.}
   { We select cluster members using different
    membership criteria, to study
   the properties of pre-main sequence stars with or without 
   circumstellar disks.}
   {We use intermediate resolution  spectra including the 
   Li\,{\scriptsize I} 6707.8\,$\AA$\ line
    to derive radial and rotational velocities, binarity    
    and to measure the Equivalent Width of the lithium line;
    these  results are combined with X-ray data to study the 
    cluster membership.
    Optical-IR data and
    H$\alpha$ spectra, these latter available for a subsample of 
    our targets,
  are used to classify CTTS and WTTS and to compare the properties 
  of stars with and without disks.}
   {We find a total of \nmembers\ certain members including 53 binaries. 
   The rotational velocity distributions of
   stars with IR excesses are statistically different from that 
   of stars without IR excesses, 
   while the fraction of binaries with disks
    is significantly smaller than that of single stars. 
    Stars with  evidence of accretion show circumstellar
    disks; youth of cluster members is confirmed by the 
     lithium abundance consistent with the initial content.}
   {As indicated by the disk-locking picture, stars with disks 
   in general have rotational velocities lower than stars without disks. 
    Binaries in NGC\,6530 seem have
    undergone significant disk evolution.}


   \maketitle
%
\section{Introduction\label{intro}} 
Very young open clusters are crucial systems for studying star 
formation mechanisms 
in different conditions
(density, metallicity, age). The present study is part of an ongoing project
 to investigate the properties of NGC\,6530, a star formation region
located in front of the giant molecular cloud M8 (Lagoon Nebula) and about 1250 pc from the Sun. 
With  stars having  a median age of 2.3\,Myr 
 \citep{pris05}, NGC\,6530  is
very suitable for investigating the stellar and disk evolution 
of pre-main sequence (PMS) stars.

Several papers have been devoted  to  this cluster 
\citep{walk57,the60,anck97,sung00,rauw02};
its low-mass population has been recently investigated by 
using X-ray data \citep{dami04},
deep {\it BVI} photometry (down to $V\sim23$, \citeauthor{pris05}, 2005)
 and 2MASS IR data \citep{dami06}.
In these recent studies,  more than 1100 candidate cluster members
have been identified
using  X-ray data and/or optical-IR photometry.

 One of the most challenging tasks in studying open clusters is to identify
certain members, especially in case of PMS stars which are located in a broad
region of the optical color-magnitude diagram. Spectra around the lithium
and H$\alpha$ lines, such as those analyzed in this work,
 are very useful to derive different and independent
membership criteria, based on the radial velocities and the lithium line.

In addition, spectroscopic observations are useful to identify 
binaries, to measure projected rotational velocities, lithium equivalent widths 
and to
distinguish stars with accretion disk phenomena by using the H$\alpha$ line.
 These
are crucial pieces of information to study the properties of PMS stars and the evolution
of their disks. 

PMS stars  with accretion
disks in which the material falls onto the stellar surface are usually
classified as CTTS.
 These objects can be spectroscopically identified since the heating of
the in-falling gas produces broad emission lines, such as 
the H$\alpha$ line. On the contrary, stars without accretion phenomena,
which show narrow chromospheric emission lines, are classified as WTTS. 

In the case of a lack of spectroscopic observations indicating
accretion phenomena, 
stars with a disk, accreting or not, can be identified using their optical-IR photometric
properties, as already discussed in several papers (see for example
\citeauthor{sici05}, \citeyear{sici05}, \citeauthor{dami06}, \citeyear{dami06}),
  circumstellar disks produce IR excesses
in the spectral energy distribution (SED) with respect to   ordinary 
 stars. Stars with IR excesses, in a small region around 
the Hourglass nebula in the M8 star-forming region, have also been  identified
by \citet{aria06}; these stars are classified as low- and 
intermediate-mass pre-main-sequence star candidates.

The excesses can be quantified using a combination
of optical-IR colors, which become useful means to distinguish stars with
or without a disk. However, while the selection of CTTS by means of the H$\alpha$
line is in general agreement with the selection of stars with disk detected
by optical-IR photometry \citep{sici05},
a clear distinction between CTTS and WTTS  cannot be obtained using only photometric 
data, since not all stars with a disk show accretion phenomena.

Circumstellar disks in PMS stars seem to be strictly related  to the angular
momentum transfer mechanisms, as recently found, for example, by
\citet{herb05,dull06,rebu06}. Therefore,
the comparison between the  rotational velocities of single PMS
stars surrounded by a disk and those of PMS stars 
which lack a disk, allows us understanding of how the
angular momentum can be regulated by circumstellar disks.
 Using a statistically large sample of
certain single cluster members and robust IR excess indicators is crucial to
test if stars surrounded by disk are slower than stars without disk,
as asserted in the disk locking model \citep{koen91}.

It is widely believed that the majority of stars form in binary systems,
but the
mechanisms of the evolution of their circumstellar disks   and their
contribution to the IR fluxes from the companion are not 
clear \citep{gras05,moni06}. Evidence of how disk properties vary
in the binary environment and with binary parameters, such as separation and 
mass, has not been found yet, since high resolution photometry and
spectroscopy have only become available in the last decade.

Another controversial question is whether the star formation process  
is instantaneous or not. Since the stars
deplete their initial lithium content in the first few million years,
   lithium abundances 
in young stars can be used as age indicators to study the age spread and hence
 the duration of the star formation process \citep{pall05}. However, accurate
 derivations of 
 lithium abundances are hampered by several difficulties related
 to the effective temperature measures and to the theoretical models, especially
  for stars cooler than 4000\,K, where the physics of the stellar atmosphere is
  not well known.

A combined study of X-ray data, optical-IR photometry and spectroscopic data 
is presented in this paper, with the aim to investigate the main properties 
of a sample of young stars in NGC\,6530, such as relations between disk presence 
and angular momentum,
accretion phenomena, binarity and lithium abundance. 
In Section\,2, we present the criteria adopted to select our targets and 
the spectroscopic
data; in Section\,3 we describe the data analysis and in particular
the technique used to derive radial and rotational velocities and
to distinguish single stars from binaries; the methods adopted 
to measure the lithium Equivalent Width
and to classify CTTS and WTTS stars based on the H$\alpha$ line
are also described. In Section\,4,
we present the analysis of stars with or without optical-IR excesses,
the veiling correction and the adopted membership criteria. Properties
related to the rotational
velocities, accretion phenomena and binarity   
of IR excess stars are presented in Section\,5; the discussion on the lithium
abundance is presented in Section\,6 and final results are summarized in Section\,7.

 
\section{Target selection and observations  \label{datared}}
\begin{figure}
\centerline{\psfig{figure=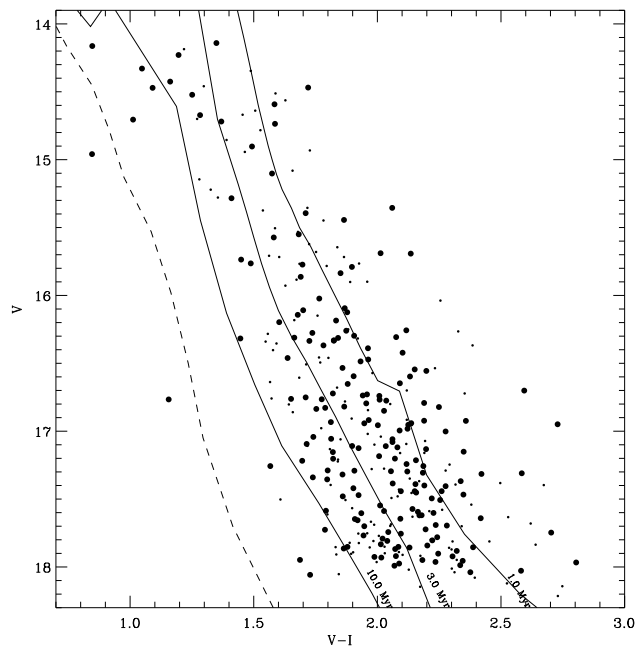,width=9cm,height=9cm}}
\caption{V vs. V-I color--magnitude diagram of the stars in the NGC\,6530
field observed with Flames.
X-ray detected targets are drawn as large  bullets.  The {\it dashed} and 
{\it solid lines} are, respectively, the solar metallicity ZAMS  
and isochrones of 1.0, 3.0 and 10.0 Myr, computed by \citet{sies00}.}
\label{5623fig1}
\end{figure}

A list of candidate members  of the star forming region NGC\,6530
suitable for spectroscopic investigations
was extracted from the optical catalog
obtained from images taken with the WFI camera of the ESO 2.2m Telescope 
and has been presented by \citet{pris05}.  As already in that
latter paper, most of the stars with an X-ray counterpart \citep{dami04} 
are located in a quite well defined region of the $V$ vs. $V-I$ color-magnitude
diagram (CMD), since most of them are cluster members. This region of the 
CMD identifies the PMS cluster locus.
We selected our targets taking stars falling in this region of the CMD,
 with $V$ magnitude between
 14.0 and 18.2.
Most of the selected targets (about 72\%) are X-ray active stars, while
the remaining ones are objects  with similar photometric properties. 
Fig.\,\ref{5623fig1} shows  the 
CMD of our targets  where X-ray detected objects are indicated 
with large bullets. The solar metallicity ZAMS ({\it dashed line}) 
and isochrones ({\it solid lines})
of 1.0, 3.0 and 10.0 Myr, computed by \citet{sies00} are also drawn, using
the \citet{keny95} transformations, the reddening value of $E(B-V)=0.35$
given by \citet{sung00} and the cluster distance d=1250\,pc derived in 
\citet{pris05}.

 A list of our selected targets is presented in Table\,\ref{5623tab1},
where 
columns\,1 to 6 give the celestial coordinates,
col.\,7 gives the spectrum number,
col.\,8 is the \citet{dami04} X-ray identification number, 
col.\,9 is the \citet{pris05} WFI identification number,
col.s\,10--15 are the \citet{pris05} $BVI$ magnitudes and their photometric
		uncertainties,
col.s\,16--21 are the IR magnitudes and photometric uncertainties\footnote{ 
'NaN'
indicates magnitudes rejected since labeled as 95\% confidence upper limits 
i.e. sources not detected or inconsistently deblended in that band.}
from the 2MASS catalog \citep{cutr03} while
col.s\,22 and 23 give  ages and masses as estimated in  \citet{pris05},
from theoretical  isochrones and tracks \citep{sies00}. 
\begin{sidewaystable*}
\begin{minipage}[t][180mm]{\textwidth}
\vspace{2.5truecm}
\tabcolsep 0.15truecm
\caption{List of stars observed with FLAMES. Sp is the spectrum number,
 ID$_{\rm X}$ is the
source identification number of \citet{dami06}; 
ID$_{\rm WFI}$ is the \citet{pris05} identification number;
$B,V,I$ and {\it eB, eV} and {\it eI}
 are the magnitudes  and their photometric uncertainties 
presented in \citet{pris05},
while $JHK$ and $eJ$, $eH$ and $eK$ are the 2MASS magnitudes and
 their photometric uncertainties;
 "NaN" indicate magnitudes rejected since labeled as 
95\% confidence upper limits
i.e. sources not detected or inconsistently deblended in that band.
 Ages and masses are those estimated in \citet{pris05}. 
The complete table is available in electronic format at CDS.} 
\centering
\begin{tabular}{ccccccccccccccccccccccc}
\hline
\hline
\\
\multicolumn{3}{c}{RA(2000)}&
\multicolumn{3}{c}{Dec(2000)}&
\multicolumn{1}{c}{Sp}&
\multicolumn{1}{c}{ID$_{\rm x}$}&
\multicolumn{1}{c}{ID$_{\rm WFI}$}&
\multicolumn{1}{c}{$B$}&
\multicolumn{1}{c}{$eB$}&
\multicolumn{1}{c}{$V$}&
\multicolumn{1}{c}{$eV$}&
\multicolumn{1}{c}{$I$}&
\multicolumn{1}{c}{$eI$}&
\multicolumn{1}{c}{$J$}&
\multicolumn{1}{c}{$eJ$}&
\multicolumn{1}{c}{$H$}&
\multicolumn{1}{c}{$eH$}&
\multicolumn{1}{c}{$K$}&
\multicolumn{1}{c}{$eK$}&
\multicolumn{1}{c}{Age}&
\multicolumn{1}{c}{Mass}\\
 h&m&s&d&m&s&&&&&&&&&&&&&&&& Myr & $M_\odot$ \\
\hline\\ 
  18& 4& 24.59& -24&19& 24.69& 2002&  493&  25079& 17.280 &  0.002 & 15.836 &  0.002 & 13.985 &  0.002 & 12.414 &  0.028 & 11.447 &  0.028 & 10.888 &  0.028 &  0.77 &  1.05\\
  18& 4& 24.62& -24&20&  4.93& 2003&  495&  23868& 17.118 &  0.004 & 15.767 &  0.001 & 14.082 &  0.002 & 12.832 &  0.032 & 12.067 &  0.032 & 11.833 &  0.032 &  1.35 &  1.38\\
  18& 4&  7.60& -24&12& 21.43& 2004&   --&  39006& 19.048 &  0.010 & 17.455 &  0.005 & 15.317 &  0.005 &    NaN &    NaN &    NaN &    NaN & 12.562 &    NaN &  1.62 &  0.67\\
  18& 4& 23.67& -24&17& 29.10& 2005&  470&  27805& 18.486 &  0.003 & 16.822 &  0.003 & 14.573 &  0.003 & 12.948 &  0.050 & 12.090 &  0.050 & 11.789 &  0.050 &  0.62 &  0.56\\
  18& 4& 16.11& -24&19& 52.26& 2006&  301&  24253& 17.227 &  0.002 & 15.790 &  0.002 & 13.893 &  0.005 & 12.373 &  0.036 & 11.449 &  0.036 & 10.829 &  0.036 &  0.64 &  0.98\\
  18& 3& 59.44& -24&12& 23.42& 2008&   70&  38914& 16.939 &  0.004 & 15.708 &  0.003 & 14.146 &  0.004 & 13.083 &  0.047 &    NaN &    NaN &    NaN &    NaN &  2.31 &  1.66\\
  18& 4& 13.38& -24&15& 13.49& 2010&  243&  32646& 18.545 &  0.004 & 16.995 &  0.005 & 14.905 &  0.002 & 13.366 &  0.058 &    NaN &    NaN & 12.246 &  0.058 &  1.27 &  0.70\\
  18& 4& 15.79& -24&21& 27.34& 2012&  295&  21339& 17.666 &  0.002 & 16.335 &  0.002 & 14.610 &  0.002 & 13.314 &  0.026 & 12.608 &  0.026 & 12.350 &  0.026 &  2.35 &  1.29\\
  18& 3& 57.28& -24&16&  9.83& 2013&   49&  30517& 17.468 &  0.003 & 16.317 &  0.002 & 14.871 &  0.003 & 13.507 &  0.072 & 12.676 &  0.072 & 12.136 &  0.072 &  9.27 &  1.38\\
  18& 4& 21.79& -24&14&  3.91& 2014&  426&  35254& 15.237 &  0.004 & 14.229 &  0.002 & 13.033 &  0.007 & 12.117 &  0.022 & 11.644 &  0.022 & 11.486 &  0.022 &  6.97 &  1.83\\
...
\\
\hline
\hline
\end{tabular}
\label{5623tab1}
\end{minipage}
\end{sidewaystable*}


The targets were observed 
with the GIRAFFE-FLAMES multi-object fiber spectrograph of the ESO-VLT Kueyen 
Telescope (UT2). We observed \nallobserved\ objects 
with the Medusa fibre system in combination with the 316 lines/mm grating
and the setup HR15 which gives a resolution  R=19300 in the wavelength range
[6606, 6905] $\AA$, comprising  the \lithiumline\ line.
Furthermore,
\nhalphasample\ of our targets were also observed using the setup HR14,
 which gives a resolution R=28800 in the wavelength range
[6383, 6626] $\AA$, comprising  the H$\alpha$ line.
These observations were made  on May 27, 2003 and
are part of the Guaranteed Time Observations of the Ital-FLAMES
Consortium.
All targets were observed using three different FLAMES fiber configurations,
as described in the Log-book in Table\,\ref{obs6530}.
Spectra taken by using the configuration ID 090229 were summed.

The data were reduced using
the  GIRAFFE Base-Line Data Reduction Software (Version  1.08)
developed at Geneva Observatory
and relative calibration files (Version 2.0).

\begin{figure}[!ht]
\centerline{\psfig{figure=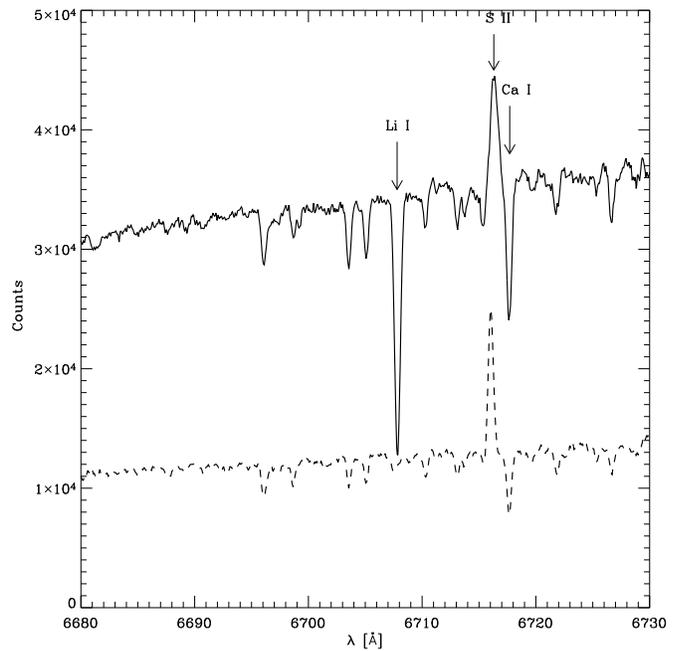,width=9cm,height=9cm}}
\caption{Examples of spectra around the \lithiumline\ line
 for an X-ray detected star,  Sp\,1083 ({\it solid line}) and for an X-ray
 undetected star,  Sp\,1052 ({\it dashed line}).
 Both spectra show several Fe {\small I} and Al {\small I} lines,
 the nebular sulphur emission feature and the Ca {\small I} 6717 $\AA$\ 
 absorption line;
 only the upper spectrum shows a very strong  Li feature.}
\label{5623fig2}
\end{figure}

Fig.\,\ref{5623fig2} shows two examples of reduced spectra around the
 \lithiumline\ line for an X--ray detected star ({\it solid line}) 
 and for an X-ray undetected star ({\it dashed line}). Both spectra show several
Fe {\small I} and Al {\small I} lines,
 the nebular sulphur emission line at $\lambda= 6717\,\AA$\ \citep{lada76}
  and the Ca {\small I} 6717.69 $\AA$\
 absorption line;
 only the upper spectrum shows a very strong  Li feature.
 
 The sky contribution from the spectra obtained around the 
 \lithiumline\ line was subtracted by using a median, 
 computed as in \citet{jeff05}, 
 of the 15 sky fiber
 spectra  available for each configuration and   
 obtained during the same night.
 The sky contribution is mainly dominated by emission of 
 the H {\small II} region,  which is spatially varying as 
 suggested  from the strong nebular emission lines which can have intensity
 variations up to 90\%. For this reason, the strong nebular emission lines
 have not been successfully subtracted, 
 while a good  subtraction of the sky continuum
 was achieved in the vicinity of lines that will be analyzed in this
 work. 
 
 Due to the strong and variable contribution of the nebular emission 
 H$\alpha$ line, we cannot subtract the sky level from our spectra
 in the H$\alpha$ region. The analysis of these spectra will be described
 in Sect.\,\ref{Halphasection}.
  
\begin{table}[!htb]
\centering
\tabcolsep 0.1truecm
\caption {Log of the observations.}
\vspace{0.5cm}
\begin{tabular}{cccccc}
\hline
\hline
\multicolumn{1}{c}{Config.}&
\multicolumn{1}{c}{Exp. Time} &
\multicolumn{1}{c}{Grating} &
\multicolumn{1}{c}{\# stars}&
\multicolumn{1}{c}{spectrum number }&
\multicolumn{1}{c}{V }\\
ID 	& [sec] &  	& 	&range	     &range\\
\hline
214312	& 5400.0&  HR15	&107	&[1002--1135]&[14.0--18.2]\\
174741	& 2808.9&  HR15 &114	&[2002--2135]&[14.0--17.7]\\
174741	& 1211.0&  HR14 &114	&[2002--2135]&[14.0--17.7]\\
090229	& 2697.1&  HR15 &111	&[3002--3135]&[14.0--18.2]\\
090229	& 3000.0&  HR15 &111	&[3002--3135]&[14.0--18.2]\\
\hline\\
\end{tabular}
\label{obs6530}
\end{table}

\section{Data analysis}
\subsection{Radial and rotational velocities and binarity \label{rvsection}}
We computed   radial velocities (RV)
 and rotational velocities of our targets using the IRAF
task {\tt FXCOR}, which performs a Fourier Cross-Correlation of each spectrum
 with a template spectrum, as described in \citet{tonr79}. We used as template
the spectrum of a relatively bright star ($V\simeq 14.5$) of our sample
(Sp 1083),
 which is regarded as a reliable
cluster member since it is a photometric candidate 
PMS star, with 
X-ray emission and a very strong lithium line (Fig.\,\ref{5623fig2} 
{\it solid line}),
 which is a feature of very young stars. 
 Among the stars with these characteristics,
this star also shows a spectrum with the narrowest lines, having
a FWHM of about 20\,km/s, which  is slightly larger than 
the instrumental spectral resolution of our spectra of about 17\,km/s,
as estimated by our 
lamp spectra;  this latter value sets a lower limit to the
rotational velocities we can determine.

For each star of our sample, we cross-correlated the spectrum with the template
over  the wavelength range between
about 6610 and 6820 $\AA$, 
 which contains no significant telluric absorption lines.
Within this  wavelength range we selected only regions without sky
or nebula emission lines.

The position of the cross-correlation peak is a measure  of the relative RV 
of each star with respect to the RV of the template star, since it is
a measure of the Doppler shift between the two cross-correlated spectra.

\begin{figure}[!ht]
\includegraphics[width=9cm]{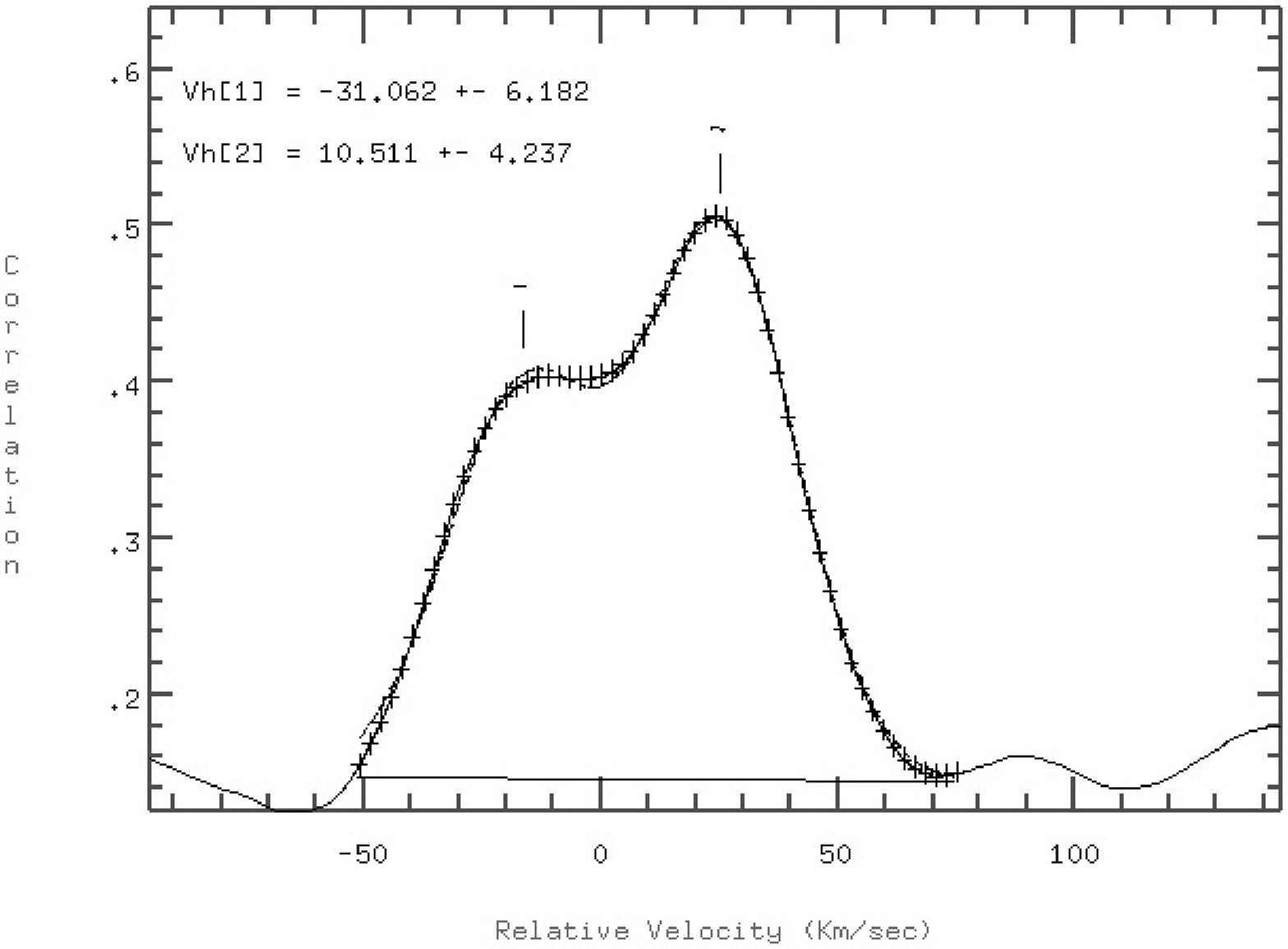}
\includegraphics[width=9cm]{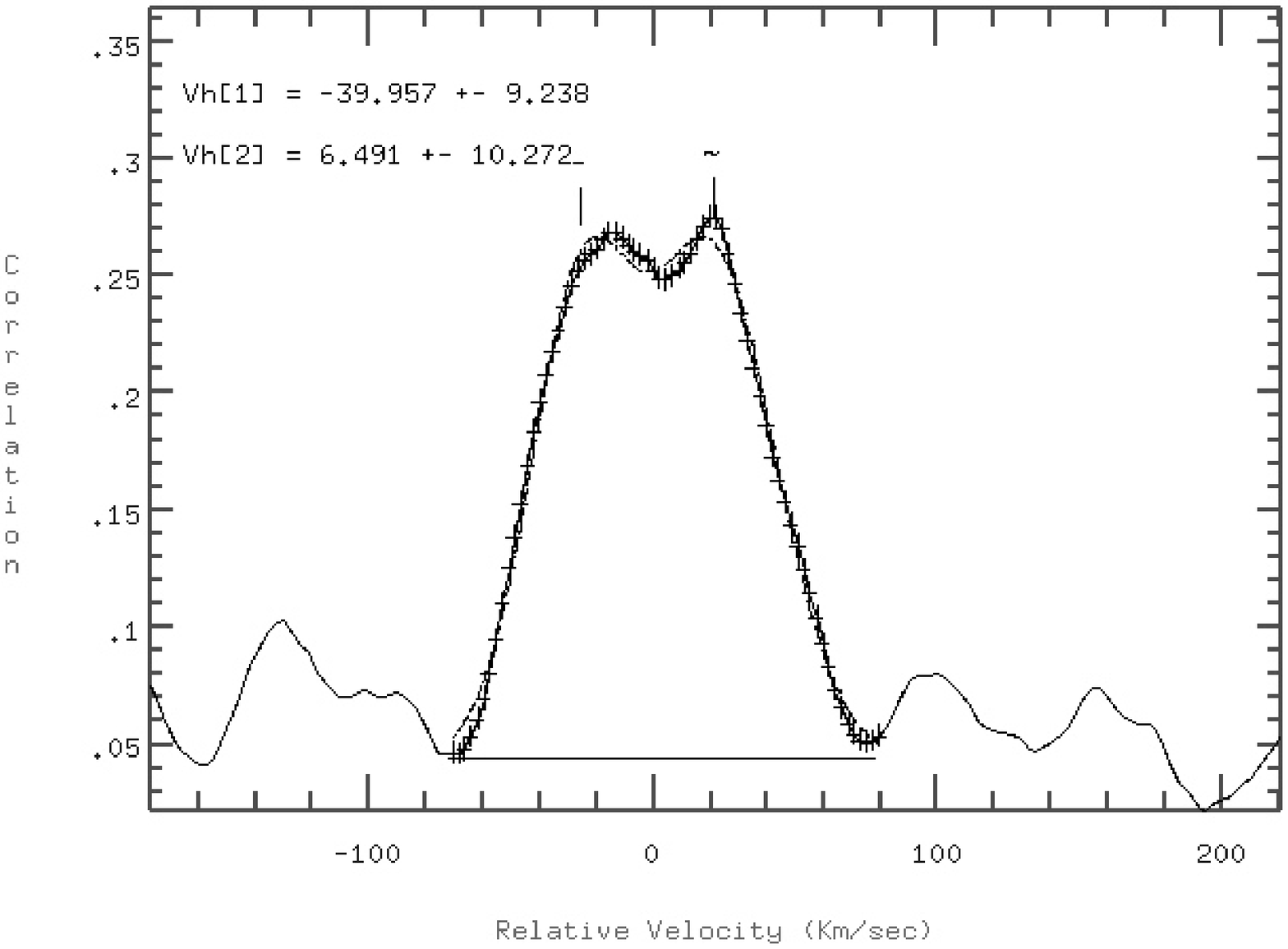}
\caption{Example of  peaks of the cross-correlation functions produced
by cross-correlating two binaries against our template star.  Relative 
radial velocities of the two components are indicated in each panel.}
\label{5623fi3}
\end{figure}

For \nbinaries\ stars 
of our sample, the peak of the cross-correlation function is not
symmetric and can be fitted with two overlapping Gaussians rather than with 
a single
one, as shown in  the two examples of Fig.\,\ref{5623fi3},  where
the relative radial velocities of the two components is also indicated.  
 In these cases, the cross-correlation function is indicative of a 
double-lined spectroscopic binary (SB2). For these objects,
however, we cannot measure the RV  of the binary systems, 
since our spectra were obtained during a single night.

To obtain the RV in the heliocentric system we proceeded as follows:
a) we used the IRAF task {\tt RVIDLINES}  to derive the observed RV
 of the star used as template,
i.e. the wavelength shift  in  spectral  lines  relative  to  rest
wavelengths;
we find that the observed RV of this star
is  $-11.85 \pm 0.56$\,km/s, based on 48 lines; 
b) using the log--book of the observations and the IRAF task {\tt RVCORRECT},
we find that the heliocentric correction to the observed RV is equal to
 $12.310\pm 0.005$\,km/s; 
c) we computed the heliocentric RV of the template star, which is equal
to  $0.47\pm 0.56$\,km/s, and from this, the heliocentric RV of all
stars (see Table\,\ref{5623tab3}). 

\begin{figure}[!ht]
\centerline{\psfig{figure=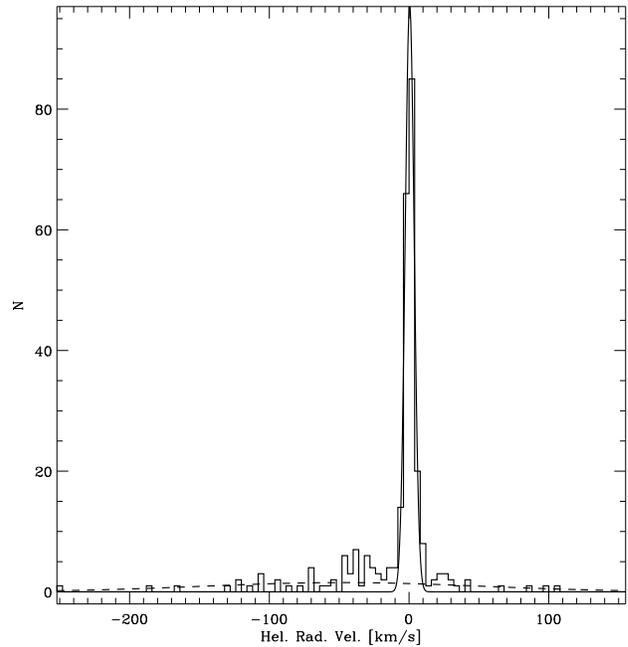,width=9cm,height=9cm}}
\caption{RV distribution ({\it solid histogram}) of the whole sample
of single stars which includes \nsingles\ objects. 
The Gaussian RV distributions for the cluster 
 ({\it solid line}) and field ({\it dashed line}),  
 obtained from the double Gaussian
fitting, are also shown.}
\label{5623fig4}
\end{figure}

The histogram in Fig.\,\ref{5623fig4} is the  RV distribution
 of the whole sample of \nsingles\ spectroscopically observed single stars; 
the histogram   shows a
significant  peak indicating  the presence of the cluster.
In order to distinguish cluster members from field stars, we fitted this
distribution with a double Gaussian using the "maximum likelihood fitting".
We find that the cluster Gaussian  is centered on 
\clusterrvmean$\pm$\erclusterrvmean\,km/s with a standard deviation  
$\sigma=$\clusterstd$\pm$\erclusterstd\,km/s ({\it solid line} in 
Fig.\,\ref{5623fig4}), 
while the much broader 
field star RV distribution shows a peak at
\fieldrvmean$\pm$\erfieldrvmean\,km/s, with a standard deviation of 
$\sigma=$\fieldstd$\pm$\erfieldstd\,km/s ({\it  dashed line}
in Fig.\,\ref{5623fig4}). These values are affected by large uncertainties
due to the poor statistics of the field star RV distribution,  reflecting
our  bias towards likely cluster members, selected from their position 
in the CMD and  the X-ray flux.

The total number of
possible cluster members within $\pm3\sigma$ of the cluster RV distribution
 is \nrvsingles\
including $\sim$\ncontaminatings\
contaminating field stars, 
as computed from the Gaussian distribution of the field stars.               

\begin{figure}[!ht]
\centerline{\psfig{figure=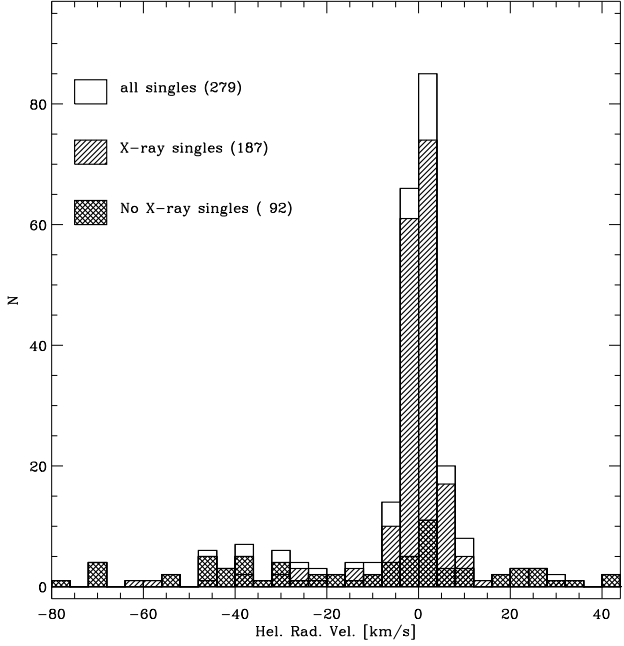,width=9cm,height=9cm}}
\caption{Zoom of RV distribution   of the whole sample
of the \nsingles\ single stars, 
compared with the  RV distribution of the \nxsingles\ single X-ray detected
stars   and with the RV distribution of the
\nnoxsingles\ single X-ray undetected
stars.}
\label{5623fig5}
\end{figure}

Fig.\,\ref{5623fig5} shows the comparison of the RV of the whole sample 
of single observed stars (\nsingles), with the RV distribution of the 
\nxsingles\ 
 X-ray detected 
 single  stars and with the RV distribution of the \nnoxsingles\
X-ray undetected
stars. The latter shows a weak peak   at the cluster RV which
 is   due to the X-ray selection incompleteness, as will be discussed in 
Section \ref{membership_section}.

The cross-correlation function allowed us to measure also the
projected rotational
velocities v\,sin(i) of our targets, since the width of the cross-correlation
peak is a function of the rotational broadening of the spectrum lines.
As discussed before,
our template is the spectrum of a slow rotation star and then 
 the width of the cross-correlation
peak with each target 
is a measure of the rotational velocity of each target with respect to the 
rotational velocity of the template.

\begin{figure}[!ht]
\centerline{\psfig{figure=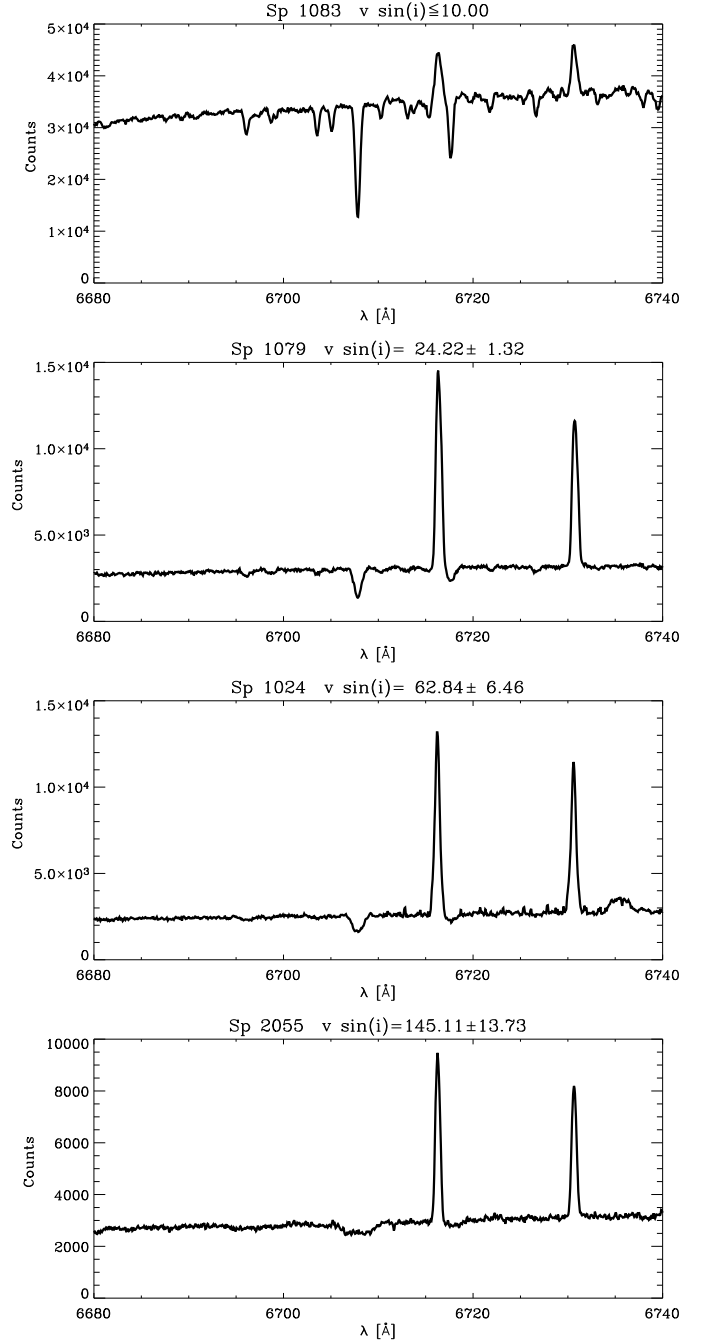,width=9cm,height=18cm}}
\caption{Examples of spectra with different broadening of the spectrum lines and
therefore with different rotational velocities. The upper spectrum is that
of the slow rotation template star.}
\label{5623fig6}
\end{figure}

Fig.\,\ref{5623fig6} shows 4 examples of spectra with different broadening
of the absorption lines and therefore different rotational velocities. 
The upper spectrum is that of the slow rotation template star.

To calibrate the relationship between the width of the cross-correlation
peak and the v\,sin(i) values, we computed synthetic spectra at different
v\,sin(i), using
the ATLAS9  \citet{kuru93} model atmospheres, assuming 
 the GIRAFFE instrumental resolution   R=19000, temperature 
T$_{\rm eff}$=4600 K (estimated from $V-I$, see
Sect.\,\ref{ewsection}),
 log $g=4.0$,  microturbulence $\xi=1$\,km/s,   macroturbulence 2\,km/s 
   and solar
metallicity, which are the expected parameters of the template star.

The obtained synthetic spectra were cross-correlated with our template and
the FWHM of the cross-correlation peak was measured. To check the sensitivity 
of this relationship on the
temperature, two additional sets of synthetic spectra were computed at 
T$_{\rm eff}$=4000\,K and   T$_{\rm eff}$=5000\,K, which are 
typical temperature
values for our targets, estimated as described in Section \ref{ewsection}.

\begin{figure}[!ht]
\centerline{\psfig{figure=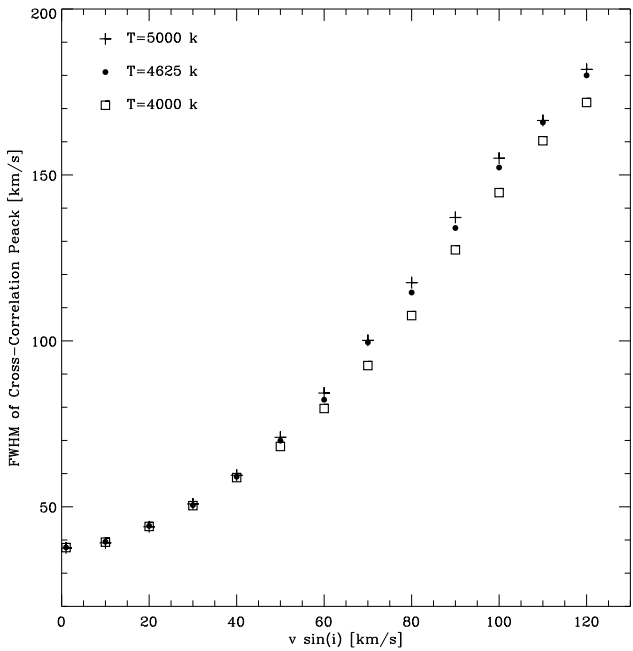,width=9cm,height=9cm}}
\caption{Calibration curve used for the v\,sin(i) estimation from the
 width of the cross-correlation peak. }
\label{5623fig7}
\end{figure}

The resulting relations between the FWHM of the cross-correlation peak and the
corresponding v\,sin(i) for the three considered temperatures are shown in
Fig.\,\ref{5623fig7}. We note that the three relations are not temperature
dependent  up to v\,sin(i)$\simeq 60$\,km/s; for v\,sin(i) $> 60$\,km/s,
the discrepancies are, however, consistent within the typical errors inferred
for the rotational velocities measured for our targets (see below in this
section).  The lower limit of the v\,sin(i) values we can measure is
approximately 
set by the instrumental resolution, equal to about 17\,km/s, which corresponds
to the flattening of the above relations of calibrations. 

Using the relationship corresponding to T$_{\rm eff}$=4600\,K, which is the
expected temperature of our template, we interpolated the FWHM of the
cross-correlation peak with the derived calibration curve to estimate the
v\,sin(i) values. Rotational velocities were derived only for single stars,
since the double peak of the cross-correlation function in SB2 stars cannot be accurately separated into the  
two components to obtain reliable v\,sin(i) values.
 As in \citet{rhod01}, the errors of the rotational velocities were taken
equal to $\pm \frac{{\rm v\,sin(i)}}{1+r}$, where $r$ is the \citet{tonr79} parameter,
which is a measure of the signal to noise ratio of the cross-correlation peak.

 Of the \nsingles\ observed single stars, \uplimitvsini\
have v\,sin(i) less than or equal to
 our estimated actual limit ($\sim 17\,$km/s) for the rotational velocities.
For \nbadspt\ stars we estimated  v\,sin(i) $\gtrsim 140$\,km/s,
but the signal to noise ratio
 of the cross-correlation peak for these stars is very low ($r\lesssim 10$)
 and  therefore it is not  possible
to distinguish whether they are  binaries and/or fast rotators.
 The large v\,sin(i) values obtained for these stars are 
 therefore to be taken with  caution.      
\begin{figure}[!ht]
\centerline{\psfig{figure=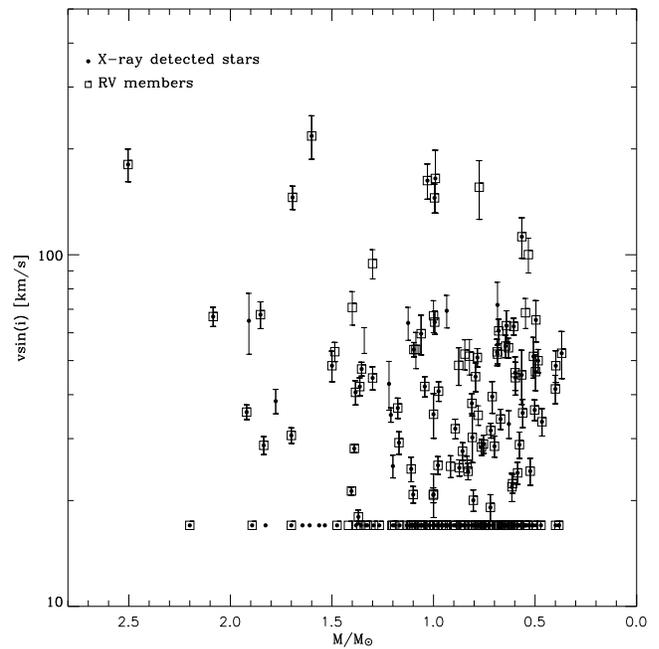,width=9cm,height=9cm}}
\caption{v\,sin(i) values vs. stellar masses for single 
stars. Cluster candidate members based on their X-ray detection and/or
RV values are indicated.}
\label{5623fig8}
\end{figure}
Fig.\,\ref{5623fig8} shows the v\,sin(i) values as a function of the 
stellar mass, for the X-ray detected stars ({\it bullets}) and for the 
RV candidate members ({\it squares}).  
The results obtained as described in this section are given in  
Table \ref{5623tab3} where
col.\,1 gives the spectrum number, 
col.\,2   indicates whether the star is SB2 (label "Y") or possible SB2 (label "?") while  
col.s\,3--6 give the heliocentric RV values and
the v\,sin(i) values,  with the estimated errors.
\begin{table*}
\centering
\tabcolsep 0.1truecm
\caption{Results of the spectral analysis obtained in this paper for the star listed in Table\,\ref{5623tab1}.
See Sections\,\ref{ewsection}, \ref{Halphasection},
\ref{IR_exc_section} and \ref{membership_section} for explanations. The complete table is available in electronic format at CDS.}
\vspace{0.5cm}
\begin{tabular}{cccccccccccccccccccc}
\hline
\hline
\\
\multicolumn{1}{c}{Sp}&
\multicolumn{1}{c}{SB2?}&
\multicolumn{1}{c}{RV}&
\multicolumn{1}{c}{$\sigma_{\rm RV}$}&
\multicolumn{1}{c}{v\,sin(i)}&
\multicolumn{1}{c}{$\sigma_{\rm v\,sin(i)}$}&
\multicolumn{1}{c}{T$_{\rm eff}$}&
\multicolumn{1}{c}{Li\,{\scriptsize I}}&
\multicolumn{1}{c}{Li\,{\scriptsize I}}&
\multicolumn{1}{c}{H$\alpha$}&
\multicolumn{1}{c}{X}&
\multicolumn{1}{c}{RV}&
\multicolumn{1}{c}{Li}&
\multicolumn{1}{c}{Q}&
\multicolumn{1}{c}{Q}&
\multicolumn{1}{c}{Q}&
\multicolumn{1}{c}{Q}&
\multicolumn{1}{c}{H$\alpha$}&
\multicolumn{1}{c}{H$\alpha$}&
\multicolumn{1}{c}{Type}\\
&&[km/s]&[km/s]&[km/s]&[km/s]&K&{\tiny EW[m$\AA$]}&{\tiny EW[m$\AA$]}\tablenotemark{\it a}&{\tiny FWZI[$\AA$]}&&&&
{\tiny \it JHHK}&{\tiny \it VIJK}&{\tiny \it VIIJ}&{\tiny \it BVJK}& class.$^b$ &PT\tablenotemark{\it c}& \\ 
\hline\\ 
 2002&   &        1.25 &        1.81 &  $\le 17.0$ &             &    4300&     532.&            --&11.3&     Y&     Y&     Y&     Y&     Y&     N&     Y&     C& PCIII&     M\\
 2003&  Y&         --  &         --  &         --  &         --  &    4500&     362.&     205.&13.4&     Y&     N&     Y&     N&     N&     N&     N&     W&    AE&     M\\
 2004&   &      -14.28 &        9.89 &        69.4 &         8.9 &    4000&      NaN&            --& 2.1&     N&     N&     N&    --&    --&    --&    --&    --&    AE&    NM\\
 2005&   &       -2.15 &       17.58 &       112.5 &        14.9 &    3900&     621.&            --&12.9&     Y&     Y&     Y&     N&     N&     N&     N&    C?&   CE?&     M\\
 2006&   &       -2.28 &        2.26 &        25.2 &         1.5 &    4200&     416.&            --&14.0&     Y&     Y&     Y&     Y&     Y&     N&     Y&     C& PCIII&     M\\
 2008&  Y&         --  &         --  &         --  &         --  &    4700&     263.&     247.&10.3&     Y&     N&     Y&    --&    --&     N&    --&    C?&   CE?&     M\\
 2010&   &        3.02 &        1.93 &  $\le 17.0$ &             &    4000&     618.&            --&10.7&     Y&     Y&     Y&    --&     N&     N&     N&     C& PCIII&     M\\
 2012&   &       -1.32 &        1.42 &  $\le 17.0$ &             &    4500&     521.&            --& 3.3&     Y&     Y&     Y&     N&     N&     N&     N&     W&    EG&     M\\
 2013&   &        1.52 &        3.76 &        40.6 &         3.2 &    4900&     393.&            --& 2.7&     Y&     Y&     Y&     Y&     Y&     N&     Y&     W&    EG&     M\\
 2014&   &      -27.04 &        1.68 &  $\le 17.0$ &             &    5800&      83.&            --&11.3&     Y&     N&     Y&     N&     N&     N&     N&     W&    AE&     M\\
...
\\
\hline
&&&&&&&&&&&&&&&&&&&\\
\multicolumn{4}{c}{$^a$EW of the second component}&&&&&&&&&&&&&\\
\multicolumn{6}{c}{$^b$C=CTTS, C?= possible CTTS and W=WTTS~~}&&&&&&&&&&&\\
\multicolumn{4}{c}{$^c$PT=H$\alpha$ profile type ({\it see text})}
\end{tabular}
\label{5623tab3}
\end{table*}

\subsection{Lithium equivalent widths \label{ewsection}}
To measure the 
equivalent width (EW)  of the lithium line of our targets,
we first  performed the continuum normalization 
in the  region around the \lithiumline\ line.
We used 
the IRAF task {\tt CONTINUUM},
considering a small line-free spectral range ($< 10 $\,$\AA$) 
around the \lithiumline\ line,
 a second order Legendre function and a variable residual
rejection limit. Due to the stellar rotation, the
FWHM of the line can vary from about 0.3\,$\AA$\ up to  2.8\,$\AA$.
 For this reason,
both the line-free spectral range and the
rejection limit were interactively chosen for each case,
based on the visual inspection
of the fitting results.

Equivalent widths of the \lithiumline\ line were measured 
with the IRAF task {\tt SPLOT};
since the EW of the lithium line is in most cases greater than 100\,m$\AA$,
we  assumed a 
Voigt profile for the majority of our targets which show a 
very strong lithium line; a Gaussian profile was considered
only in the few cases where the EW of the line 
was smaller than about 100\,$\AA$.
 The EWs were not corrected for the contribution due to the 
nearby Fe {\footnotesize I}  line at 6707.44$\,\AA$ 
blended with the Li line, since for our targets
this contribution is negligible with respect
to the lithium. Following the method given in \citet{sode93}, 
for the stars of our sample
the EW of the Fe
can range from 10 to 20\,m$\AA$,  which implies a
systematic contribution of a few percent for our EW measures, that are 
typically larger
than 200\,m$\AA$.
   
In order to estimate uncertainties in the EW measures, we repeated 
four times both
the continuum normalization and the EW measures. 
The maximum error distribution of such measures does not 
depend on the S/N nor on the FWHM, but slightly increases with
the mean value, up to a value of about 0.15\,$\AA$\ for EW$\sim 0.6$\,$\AA$.
We then considered as final EW values the mean of the four measures and 
as typical error   the standard
deviation of the maximum errors of all measures,
 which is 0.035\,$\AA$ for all EW. 

Values of the lithium EW for SB2 stars were computed with respect 
to the combined continuum which is 
the sum of the continuum contributions of both components. 
Lacking information on the individual spectral types of the two components, we
cannot separate their continuum contributions and therefore
we cannot obtain individual estimates of the lithium EW for these stars.

 The EW of the lithium line in young stars depends strongly  on 
the star temperature.
 Lacking information of spectral types,
accurate temperatures  of our targets are not available. For this reason,
temperature estimates have been derived from photometry as follows:
we dereddened  observed $(V-I)$ colors   by subtracting the mean reddening
$E(V-I)=0.46$, obtained from the $E(B-V)=0.35$ \citep{sung00}
and the reddening law $E(V-I)=1.3E(B-V)=0.35$ \citep{muna96}.
As already done in \citet{pris05}, the dereddened $(V-I)_0$
colors were converted into effective temperatures using the relation of
\citet{keny95}. 

\begin{figure}[!ht]
\centering
\includegraphics[width=9cm]{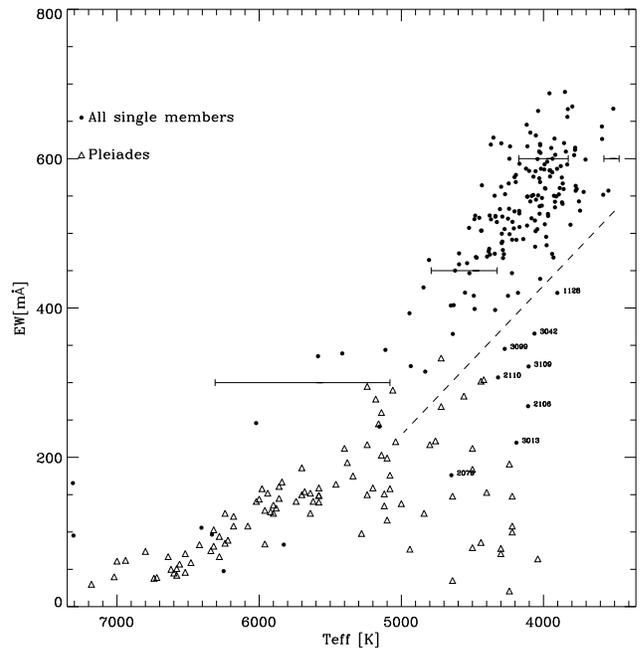}
\caption{Lithium EW for candidate single members of NGC\,6530. Lithium EW
of the older Pleiades stars are also shown for comparison. {\it Horizontal bars}
indicate expected temperature errors, while the {\it dashed line} indicates the
limit adopted to define candidate Li depleted stars. The 8 stars with evidence
 of Li depletion and/or veiling effects are also indicated.}
\label{5623fig9}
\end{figure}

The errors on effective temperatures
 strongly depend
on the unknown individual reddening, which is the largest source of error.
To estimate temperature errors, we considered four typical observed $V-I$ values
 (1.25, 1.70, 2.15, 2.80) which we dereddened  using the  
 reddening range
$E(B-V)=[0.25,0.50]$,  found by \citet{sung00}. Typical temperature
errors are indicated as horizontal bars in Fig.\,\ref{5623fig9}, where 
the lithium EWs of all single members are shown  
as a function  of temperature. 

In order to verify the young age of our targets,
the lithium EWs of older stars in the Pleiades
cluster (100 Myr) were also plotted \citep{sode93}.
 The comparison of the two samples of stars
clearly shows that the EW of most of our targets  is significantly 
higher than those
of stars of similar temperature in the Pleiades. 
This is consistent with theoretical
models that predict a cosmic abundance of lithium in stars younger than
about 10\,Myr \citep{dant97,sies00}.

EW estimates of the lithium line for our targets are reported in 
 Table\,\ref{5623tab3}, where 
col.\,7 gives the temperature of the stars, 
col.\,8 gives the EW of the lithium for single stars
or for the first component, if the object is identified as SB2; in these latter cases,
 the lithium EW of the second component is given in col.\,9.
The label "NaN" indicates that either the line is not present or it cannot be
 measured.

\subsection{H$\alpha$ observations \label{Halphasection}}
\begin{figure}[!ht]
\centerline{\psfig{figure=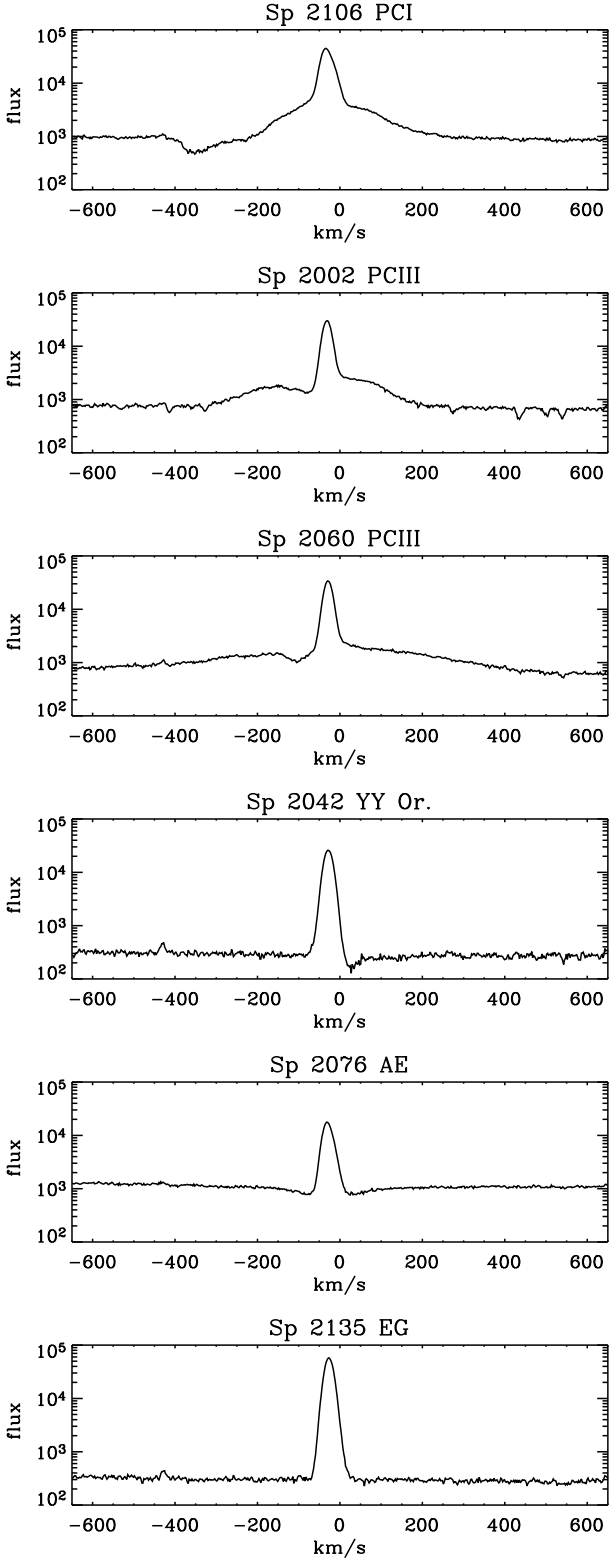}}
\caption{
Examples of H$\alpha$ line profile as a function of the Doppler velocity.
Note that fluxes are shown on a log scale.}
\label{5623fi10}
\end{figure}
All our targets are projected onto  the Lagoon Nebula, a very bright
HII region, which produces a strong and  
 spatially non uniform emission line background.
This means that all spectra of our targets include a contribution of 
a spatially variable H$\alpha$ emission, due to the Nebula itself.
Since this contribution cannot be individually measured and 
subtracted from the stellar spectra,
we are not able to measure the purely stellar EW of the H$\alpha$ line.

By inspection of the H$\alpha$ spectral region, however, it is obvious 
that a number of stars show broad lines, while the  H$\alpha$ from the Nebula
is very narrow. This allows us to identify these latter stars as CTTS, since
their H$\alpha$ line indicates gas motion in extended envelopes. 

To select stars with signatures of accretion 
we measured the Full Width at Zero Intensity (FWZI)
of the H$\alpha$ line for the subsample of  \nhalphasample\
stars for which the spectrum around the H$\alpha$ region is available.
The measures are reported in Col.\,10 of 
Table\,\ref{5623tab3}; the errors  depend on the
signal to noise ratio and are of the order of a few $\AA$.

Different kinds of H$\alpha$ line profiles are found
and classified following the notation described in \citet{bert84} 
(see col.\,19 of Table\,\ref{5623tab3});  
some examples are reported in Fig.\,\ref{5623fi10}. 
 As summarized in the tree diagram shown in Fig.\,\ref{treedgr},
of all the 95 single stars for which  the H$\alpha$ spectrum
is available, 71 are certain members while 2 are possible members.
Among the certain members,
31 show signatures of accretion and we classify them as CTTS; 
in particular we have:
1 star (Sp 2106) showing a type I P Cygni profile (label 'PCI'); 
28 stars showing from one to three broadened emission components, 
as for example 
Sp 2002 and 2060 of Fig.\,\ref{5623fi10}; these objects are classified as 
type III P Cygni profile (label 'PCIII');
finally, 2 stars have inverse  P Cygni profiles, as Sp 2042 (label 'YY Ori').
In addition, we have
9 stars with slight broad wings which can be due to "normal" chromospheric
emission or to low accretion; we label these objects as "CE?" and we classify
them as possible CTTS or "CTTS?".
On the other hand, we have a total of 31 member stars   
which do not show evidence of accretion:
6 of them show a broad absorption H$\alpha$ line,
 with the core filled of the narrow nebular emission H$\alpha$, that
we label as 'AE' (Sp 2076 is an example), while the remaining
25  stars show a quite symmetric and "narrow" 
line profile, likely originating only from the nebular emission; 
we label these
objects as 'EG' and an example is the star Sp 2135. We classify these
31 objects as WTTS.    The 2 possible members show a H$\alpha$ line profile
labeled as 'AE'.
  
These  results are also reported in Table\,\ref{5623tab3}
where col.\,18   indicates
if they are classified as CTTS (label 'C') or possible CTTS (label 'C?') 
or   WTTS (label 'W'); col.\,19 indicates the profile type of the 
 H$\alpha$ line.
\begin{figure*}[!ht]
\caption{ Tree diagram showing the classification of the 95 single stars
with different kinds
of H$\alpha$ profiles. Labels are described in Sect.\,\ref{Halphasection}.}
{\scriptsize
\Tree[.{95 singles \\ w/ H$\alpha$} [.{71 M} [.{31 \\ CTTS} {1 PCI} \
{28 PCIII} {2 inv PC} ] {9 CE? \\ CTTS?} [.{31 WTTS} {6 AE} {25 EG} ] ] {2 M?} {22 NM} ]}
\label{treedgr}
\end{figure*}
%

\section{Analysis}
\subsection{Stars with infrared excesses \label{IR_exc_section}}
 IR excesses are indicators of presence of circumstellar disks, typical
of T Tauri stars.  A study of IR-excess stars in NGC\,6530 \citep{dami06}  
has shown that different optical-IR excess indices are related to 
different kinds of 
spectral energy distributions.

We use, here, the extinction-free indices
\begin{eqnarray} 
Q{\rm _{JHHK}=(J-H)-E(J-H)/E(H-K)\times(H-K)},\nonumber\\
Q{\rm _{VIJK}=(J-K)-E(J-K)/E(V-I)\times(V-I)},\\
Q{\rm _{VIIJ}=(I-J)-E(I-J)/E(V-I)\times(V-I)},\nonumber\\
Q{\rm _{BVJK}=(J-K)-E(J-K)/E(B-V)\times(B-V)},\nonumber
\end{eqnarray}
already 
defined in \citet{dami06}, to select, in our sample, stars with IR excesses
in at least one of the four indices.  
The \citet{riek85} interstellar extinction
 law
has been used for the JHK magnitudes.
We consider only objects with an IR magnitude 
error smaller than 0.1 mag. 
To avoid non-member
contaminating stars we consider only objects that are members 
according to at least two of our 
membership criteria,  based on X-ray detection, RV and presence 
of the lithium line (see Section\,\ref{membership_section}). 

\begin{figure*}[!ht]
\centerline{\psfig{figure=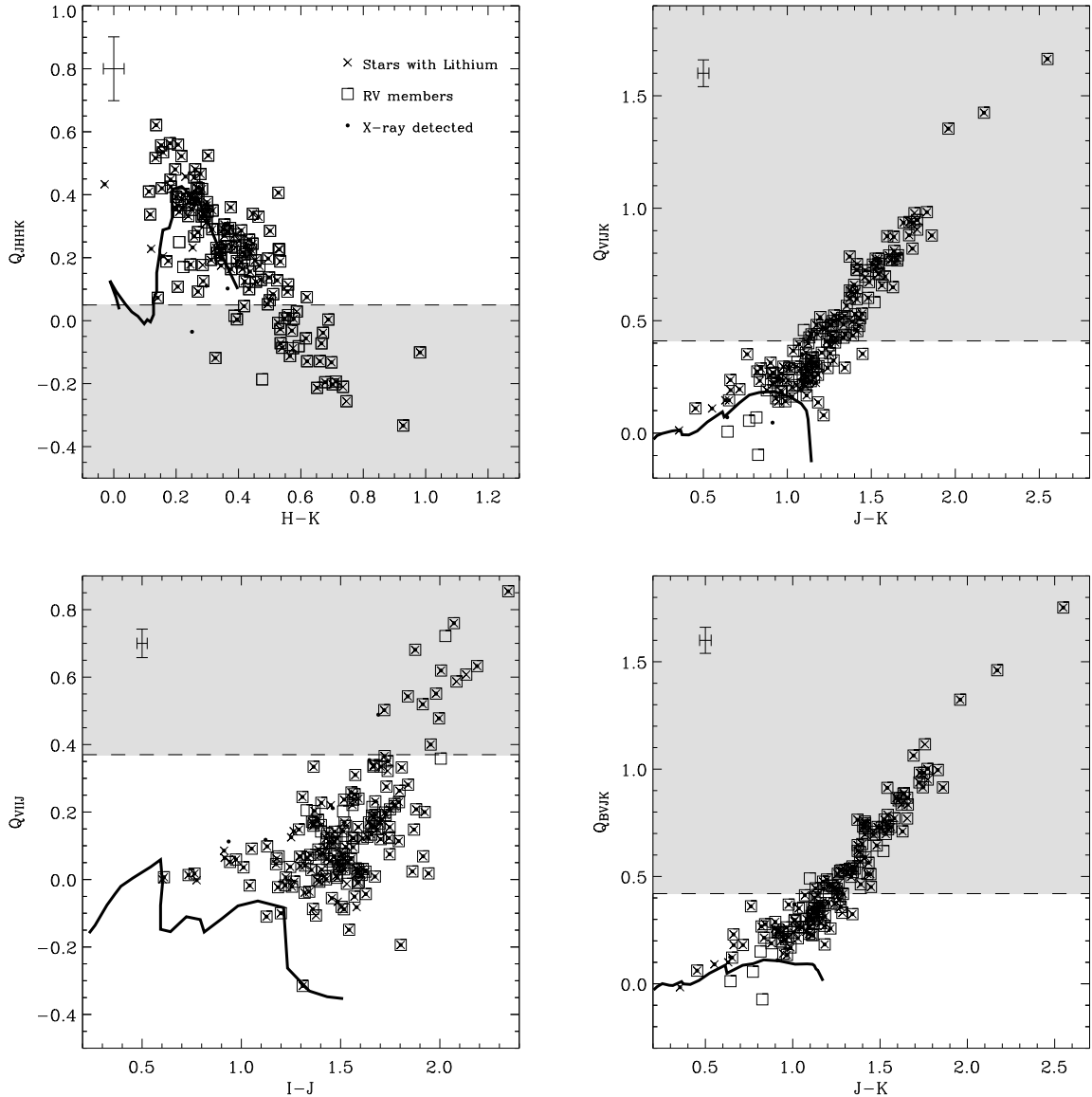}}
\caption{Extinction-free indices defined in \citet{dami06}, as a function
of colors. Different symbols indicate cluster members selected
using different membership criteria. The {\it solid line} indicates normal star
colors from \citet{keny95}, while the {\it dashed lines} indicate the limits
 used to define the regions with IR excess stars 
 ({\it gray shade regions}). 
 Typical error bars in both directions
 are plotted in each panel.} 
\label{5623fi12}
\end{figure*}

Fig.\,\ref{5623fi12} shows our computed
optical-IR indices as a function of one optical or IR color, for our candidate
single members; different symbols indicate the membership criterion. 
 Typical propagated errors are plotted in each panel, where 
the {\it solid line} indicates normal star
colors from \citet{keny95}, 
for $E_{B-V}=0.5$, which
is the maximum reddening estimated for the cluster. 

 We define "IR excess stars",
the  objects falling in the gray shade regions 
of Fig.\,\ref{5623fi12}, i.e. stars 
with  indices exceeding 
 the fiducial limits
defined in \citet{dami06}
and indicated in each panel by the {\it dashed line}. 
 The selection has been made
regardless of  photometric error. The stars 
that we have classified as certain members 
(see Section \ref{membership_section})
having IR excesses with 1-$\sigma$ error
of the index compatible with normal colors,
are 13 from a total of 81. Nevertheless, only a fraction of
these 13 stars could not be stars with IR excesses.

The number of stars with each of our membership indicators that are also disk
candidates based on each of the IR excess indices 
is given in Table\,\ref{5623tab4}. 
For each membership indicator (X, RV, Li), the last two lines of
Table\,\ref{5623tab4} give also
the number of stars showing IR excess based on at least one of the four
indices (Q$_{\rm xxxx}$) 
and the total number of stars with the given membership indicator.

Stars with and without IR excesses are reported in Table\,\ref{5623tab3} 
where col.s\,11--13  indicate whether the star has an X-ray counterpart and/or if
it is member based on the RV and/or if the lithium line is present while
col.s\,14--17 indicate whether the star shows  IR
excess in the corresponding Q index.
Properties of the stars with or without IR excesses are
discussed in Section\,\ref{prop_IR_section}.
\begin{table}
\centering
\tabcolsep 0.2truecm
\caption{ Number of stars with each of our member indicators
(X-ray detection, RV  measurements and EW of the Lithium line),
which are also disk candidates based on the IR excess indices.
For each membership indicator (X, RV, Li), the last two lines
give also the number of stars showing IR excess based on at
least one of the four indices (Q$_{\rm xxxx}$)
and the total number of stars with the given
membership indicator.}
\begin{tabular}{|l||*{4}{c|}}\hline
\makebox[3em]{}
&\makebox[3em]{X}&\makebox[3em]{RV}&\makebox[3em]{Li}\\\hline\hline
Q$_{\rm JHHK}$&          26&          32&          29\\\hline
Q$_{\rm VIJK}$&          67&          76&          74\\\hline
Q$_{\rm VIIJ}$&          12&          14&          13\\\hline
Q$_{\rm BVJK}$&          68&          76&          74\\\hline
Q$_{\rm xxxx}$&          76&          84&          81\\\hline
Tot. &         187&         190&         195\\\hline
\end{tabular}
\label{5623tab4}
\end{table}

\subsection{Veiling}
As in \citet{pall05}, we estimated the amount of spectral veiling
affecting our stars, using the three strong lines of
V\,{\footnotesize I} at 6624.8\,$\AA$,
Ni\,{\footnotesize I} at 6643.6\,$\AA$\ and 
Fe\,{\footnotesize I} at 6663.4\,$\AA$. We cannot compare the EW of these lines
in our targets with those measured in stars of IC\,2391 and IC\,2602 
\citep{rand01} and
used by \citet{pall05}, since the spectral resolution
is  $R\sim 43\,800$,  significantly higher than that of our spectra
($R\sim 19\,300$). 
To give a consistent estimate of veiling in our targets, we 
compared the EW of the selected lines with those of WTTS candidates
selected in our sample.
 WTTS candidates  were chosen as  
those objects that are X-ray detected, members based on the RV,
showing the lithium line,
and without IR excesses with respect to the $Q$ reddening-free 
indices defined in Section\,\ref{IR_exc_section}.
This condition ensures that we select stars without strong disks and likely
not affected by veiling. To avoid large errors
due to rotational broadening  in the 
EW measures of the three lines, we considered only slow rotation members 
(v\,sin(i) $<$ 30\,km/s).

\begin{figure*}[!ht]
\centerline{\psfig{figure=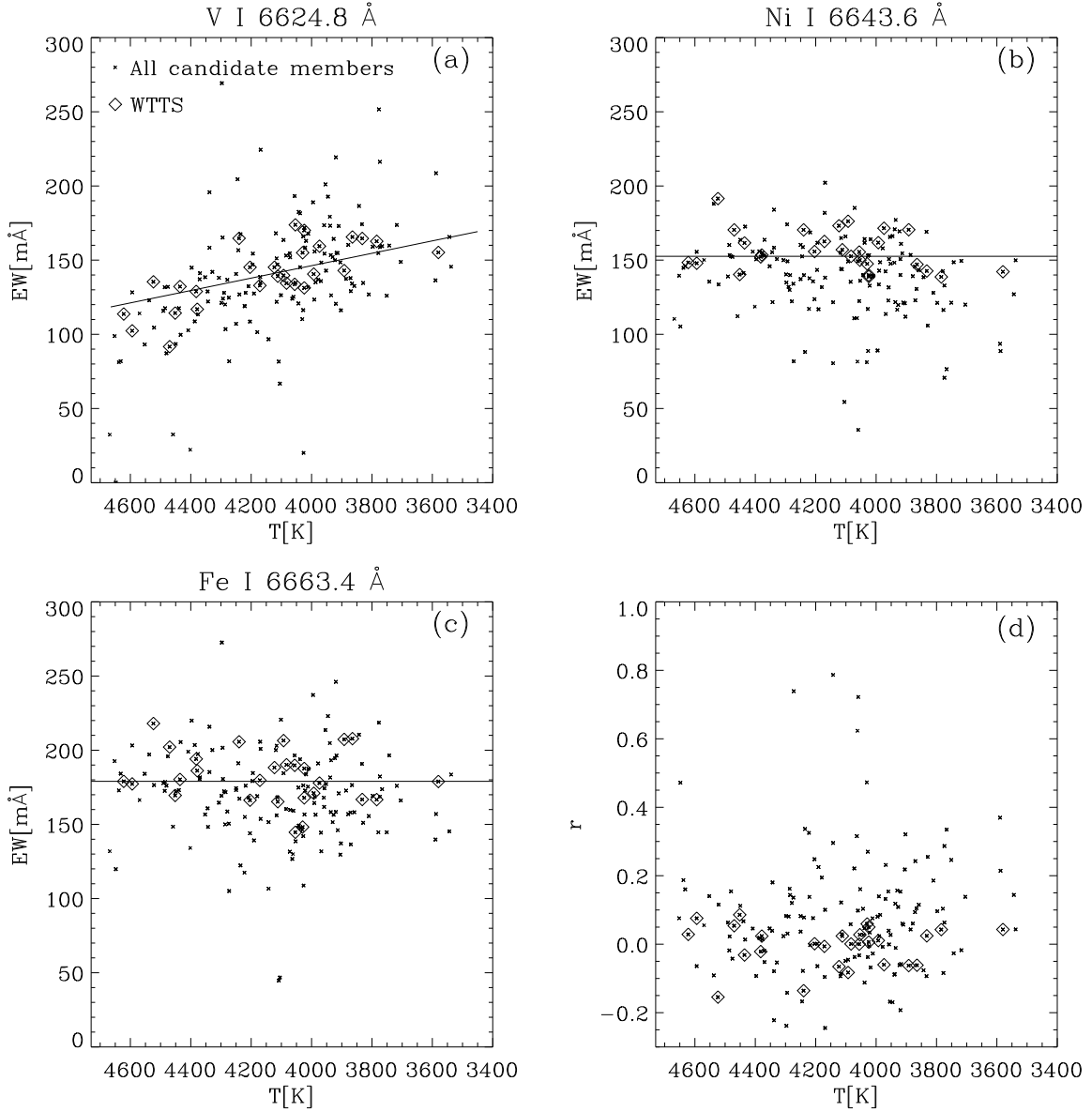}}
\caption{{\it Panels} (a), (b) and (c) show the EW as a function of the
temperature of the V\,{\scriptsize I}, Ni\,{\scriptsize I} and 
Fe\,{\scriptsize  I} lines
used to estimate the veiling amount in our targets. The average values of the 
estimated veiling r as a function of 
temperature are shown in {\it panel} (d).}
\label{5623fi13}
\end{figure*}

Panels $a$, $b$ and $c$ of  Fig.\,\ref{5623fi13}
show the EWs of the three lines measured for all
our targets that are members according to at least one criterion.
Our WTTS candidates
are also indicated. We do not have a significant number of
WTTS for temperatures higher than $\sim$4700\,K and therefore we 
can estimate the veiling contribution only for stars cooler than this
temperature.   

For each of the three lines 
and each candidate, we computed the veiling correction as  
${\rm r}={\rm EW}_{\rm WTTS}/{\rm EW}_{\rm stars}-1$ 
where ${\rm EW}_{\rm WTTS}$ has been obtained  from a linear
fitting  of the EW of the WTTS, for the V\,{\footnotesize I} line, while
it has been taken as the median value of the  EW of the WTTS, 
for the Ni\,{\footnotesize I} line and the Fe\,{\footnotesize I} line,
for which no trend of the EW as a function of temperature is found.
Panel $d$ shows the average
r values of the three lines, where the dispersion is about 0.3. 

For the bulk of our stars, 
these values of r are consistent with zero or a small correction
to the measured Li EW
(considering the dispersion as the typical error). In addition, the
spread of the r values is comparable to the typical lithium EW error.  
 
We checked that r is not significantly correlated 
with either the IR excess
indices or the FWZI of H$\alpha$.  
This suggests that our veiling determination is affected by large errors,
and so would be for any correction to the Li EWs based on it. Such a correction
would have an error comparable to EW measurement errors themselves. 
We have applied this correction only 
for a few possible members with Li EW smaller 
than that in stars of similar temperature (see Section\,\ref{alisection}).
%
\subsection{Cluster membership criteria \label{membership_section}}
Our data allow us to assign the cluster membership to our targets, 
 based on
the combination of RV measurements, X-ray detection and EW of the lithium
line. In addition, the profile type of the H$\alpha$ emission,
when available, and optical-IR excesses, have been used as
 further indications of cluster membership.
 
Of all \nallobserved\ spectroscopically observed stars, we find
\nsingles\ single and \nbinaries\ binary stars; of them
\nxsingles\ singles plus \nxbinaries\ binaries 
are X-ray detected, \nrvsingles\ single stars have
 RV   consistent within $3\sigma$ with the cluster mean RV,
  and \nlisingles\ singles
plus \nbinaries\ binaries show a measurable lithium line. We consider  the 
\nrvandxandlisingles\ X-ray detected single stars showing the lithium line,
with a RV consistent with that of the cluster as certain cluster members.
In addition, we have \ndepleted\
cases for which the Li EW  is smaller than that 
in stars of similar temperature. We include them in the sample of certain 
cluster members; a discussion about these stars will be 
presented in Section\,\ref{alisection}. 
 
To the sample of cluster members we add  \nsbuno\ stars that are X-ray
detected with a RV inconsistent with cluster
membership, but Li EW larger than a temperature dependent threshold value 
(see Section \,\ref{alisection} and Fig\,\ref{5623fig9}) and
 consistent with that of certain cluster members.  
We assume that these are binaries, members of the cluster,
with the spectrum of only one of the two components being visible (SB1). 
We also consider as possible cluster members the
 \nxmembdepleted\ X-ray
detected stars with RV inconsistent with cluster
membership and Li EW smaller than the adopted threshold limit for the lithium EW;
we labeled these stars with 'M?'. Nevertheless, we cannot rule out that these are                                               
 young field stars, unrelated to and slightly older than NGC\,6530 stars.

On the other hand, we have  \nrvlinoxsingles\ additional single stars that are
members based on the RV,
  with the lithium EW larger than the adopted threshold limit, 
but are not detected in X-ray. This is consistent
with the expected incompleteness of our X-ray selected sample. 
All these stars have mass lower than 1.5\,$M_\odot$ ($V>16.3$, 
T$_{\rm eff}<$4800\,K); 
we classify them as certain
members ('M') while we consider as possible cluster members ('M?') the
\nrvmembdepleted\ stars that are X-ray undetected, with a RV consistent with the cluster members but with
the lithium EW smaller than those of similar temperature stars;
non members are indicated as 'NM'. Final membership results are reported in
col.\,20 of Table\,\ref{5623tab3}.

The number of stars found with the adopted membership criteria
are given in Table\,\ref{5623tab5}.
 \begin{table}
\centering
\tabcolsep 0.1truecm
\caption{Number of single X-ray detected and undetected stars
({\it bottom Table}) with different properties: RV members/non-members (Y/N)
within $3\sigma$, Li test certain members (Y), possible members (Y?) and non-members (N) (see text).
The number of certain
members are reported in  boldface and those of possible members are in italic.}
\vspace{0.5cm}
\begin{tabular}{l*{1}{c}}
\makebox[18em]{Single X-ray detected}\\
\end{tabular}
\begin{tabular}{|l||*{4}{c|}}\hline
\backslashbox{RV}{Li}
&\makebox[3em]{Y}&\makebox[3em]{Y?}&\makebox[3em]{N}&\makebox[3em]{Any}\\\hline\hline
Y	&{\bf         158}&{\bf           8}&           0&         166\\\hline
N     &{\bf           8}&{\it           7}&           6&          21\\\hline
Any   &         166&          15&           6&         187\\\hline
\end{tabular}

\vspace{0.5cm}
\begin{tabular}{l*{1}{c}}
\makebox[18em]{Single X-ray undetected}\\
\end{tabular}
\begin{tabular}{|l||*{4}{c|}}\hline
\backslashbox{RV}{Li}
&\makebox[3em]{Y}&\makebox[3em]{Y?}&\makebox[3em]{N}&\makebox[3em]{Any}\\\hline\hline
Y	&{\bf          10}&	{\it           3}	&          11&          24\\\hline
N     &           0&           1&          67&          68\\\hline
Any   &          10&           4&          78&          92\\\hline
\end{tabular}
\label{5623tab5}
\end{table}

We have a total sample
 of  \nmembersingles\ (158+8+8+10) 
 single certain members (indicated in the table with boldface characters),
 where we also included the \ndepleted\ candidate depleted certain members
 (see Section\,\ref{alisection});
 by adding the \nbinaries\ 
 binaries showing
strong lithium\footnote{all \nbinaries\, binaries except one are X-ray detected}
 we have a total sample
 of \nmembers\ certain members, plus \npossmembersingles\ (7+3)
 possible single cluster members (indicated in the table with italic characters);
  thus, about 74\% 
 of our initial sample are certain and possible cluster members.
 
  Our data also allow us to estimate the  
completeness of the X-ray data.  By considering the sample of single
stars, the X-ray completeness  has been estimated as the ratio between the
number of X-ray detected stars, which are also members according to
the lithium  test and
the total number of candidate members from Li, but  regardless of X-rays.

The fraction of X-ray detected stars in ACIS FOV and cluster locus, which were
observed with GIRAFFE/FLAMES (setup HR15), is 0.61 (in the range
V=14-18.2).
The corresponding fraction for X-ray undetected stars (within the same
magnitude and color range) is 0.25.
Therefore, the inverse of these numbers (1.64 and 4.00, respectively)
are the corrections for completeness, 
which we need to apply to our FLAMES samples
to estimate the
actual number of members (either X-ray detected or not) in the cluster
locus down to V=18.2, which thus become 
(181*1.64)=296.66 (X-ray detected members) and
(181*1.64)+(14*4.00)=352.66 (all members), respectively,
if we include in these samples certain (M) and  possible members (M?). 
Taking into account the contaminating members (6), 
the X-ray completeness is equal to 85\% while,
 if we consider only certain members, the 
X-ray completeness is equal to 
89\%.
The comparison of the two X-ray completeness values 
allows us to estimate an error of the order of 4\% in the X-ray completeness estimate.
These values are consistent with the adopted X-ray incompleteness correction
applied to the Initial Mass Function in \citet{pris05}. 

\section{Properties of IR excess stars: rotational velocities, accretion and 
binarity \label{prop_IR_section}}
\subsection{ Introduction}
IR excesses are useful to discriminate PMS stars without disks (WTTS)
 from those with disks, which can also have  accretion of material 
onto the star. On the other hand, T Tauri stars of similar spectral type
show a trend with rotational velocity, in the sense that
stars surrounded by accretion disks  seem to be slow rotators. 
This has been explained in terms of disk locking \citep{koen91}, but observational
evidence for a correlation between rotation periods and accretion disk indicators has 
been debated \citep{edwa93,stas99,herb00}.  
In  \citet{rhod01}, the 
comparison of the v\,sin(i)
distributions of CTTS and WTTS in the Orion Nebula Cluster, 
selected by means of $I-K$  excesses and Ca\,{\small II}
emission line strength, suggests, although the result is not 
statistically significant,
 that the CTTS sample contains a slightly larger
fraction of slow rotators than the WTTS sample. 
On the contrary, using a different CTTS versus WTTS classification 
based on the H$\alpha$ line of the same
sample of stars, \citet{sici05} found a significant difference  between the
two distributions. More recently, using Spitzer data, a clear correlation
between rotation and IR excesses  has been found in Orion 
by \citet{rebu06}. 

By using extinction-free indices Q$_{\rm ABCD}$ (with ABCD   generic IR-optical
magnitudes)
as indicators of the presence of disks,
 we selected a sample of cluster members
with IR excesses,
as described in Section\,\ref{IR_exc_section}. On the other hand,
we defined  "No IR excess WTTS" as those members that do not show  IR excesses in
any of the considered indices Q. This allows us to select stars without 
strong disks
and therefore, without strong signatures of accretion. In order to exclude
high mass fast rotators, we considered only K and M
type stars,  with a temperature between 3700\,K and 4500\,K,  estimated 
based on the $V-I$ colors.

\subsection{ Rotational velocities}
\begin{figure}[!ht]
\centerline{\psfig{figure=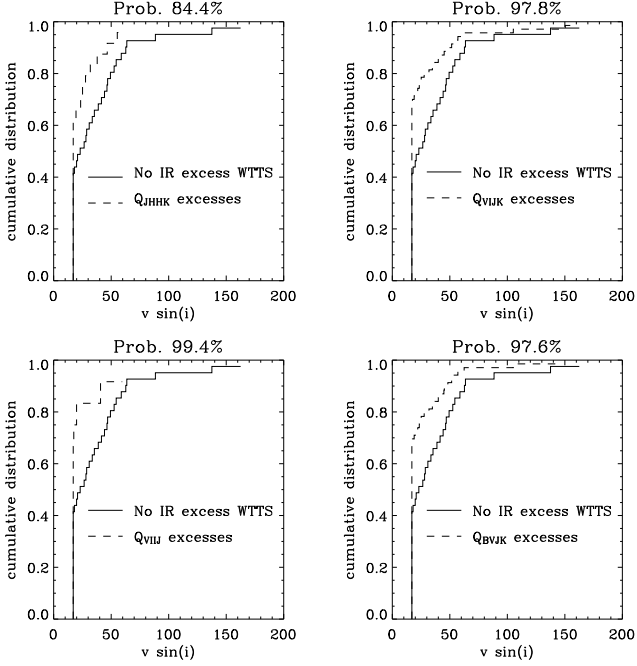,width=9cm,height=9cm}}
\caption{Cumulative distributions of the v\,sin(i) values for the sample of the
WTTS stars and for the samples of stars with IR excesses in the corresponding
indices. The probability that the two distributions are significantly
different is given in each panel.}
\label{5623fi14}
\end{figure}

To quantify possible effects on the rotational
velocities
due to the circumstellar disks,
we compared the distribution of the  v\,sin(i)
of  stars with IR excesses with the  distribution of the  v\,sin(i) of the
candidate WTTS stars, as shown in Fig.\,\ref{5623fi14}.  K-S tests have been
made to compare the  v\,sin(i) distribution of the "No IR excess WTTS" 
and those of IR excesses stars. The resulting probabilities that the two distributions
 are distinguishable 
(shown on the top of each panel) indicates that
the distributions are statistically different at a significance level
$ \sim 98\%$ for the Q$_{\rm VIJK}$ and  Q$_{\rm BVJK}$ excess stars
 and $ \sim 99\%$ for the Q$_{\rm VIIJ}$ excess stars. 
This  result allows us to support previous observational
evidence that stars with disks (which can be CTTS stars) are slower rotators
than "No IR excess WTTS",
as indicated in the disk locking picture\footnote{As already discussed,
 IR excesses are signatures of circumstellar
material around the central star. However, not all stars with a disk
show accretion phenomena and then our adopted indices  cannot help
us to distinguish CTTS  with accretion disks
from stars with a disk
but without accretion.}
 and as already found
in the Orion Nebula Cluster \citep{herb02,sici05,rebu06}.
\begin{figure}[!ht]
\centerline{\psfig{figure=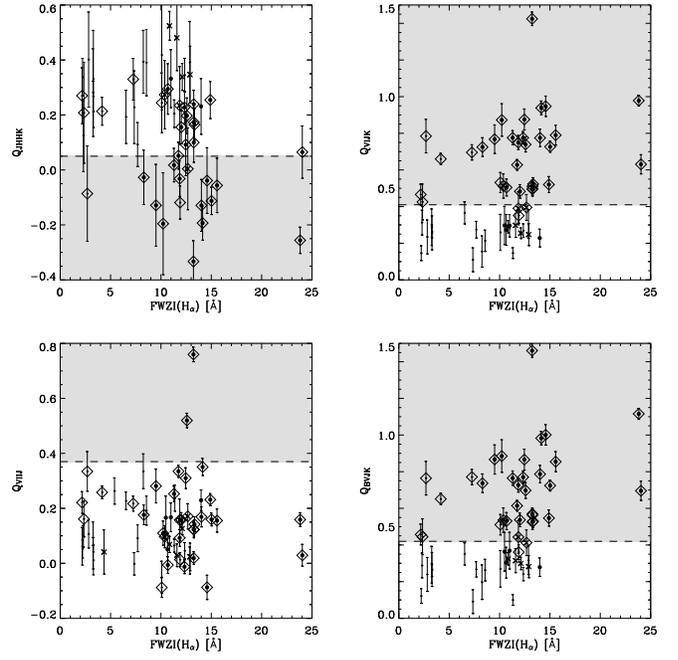,width=9cm,height=9cm}}
\caption{Extinction-free indices of IR excesses vs. the H$\alpha$  FWZI
for the sample of stars for which the spectra in the
 H$\alpha$ region are available. {\it Small dots} and {\it large bullets} 
 are WTTS and CTTS, respectively,
defined based on the H$\alpha$ line, 
{\sf x} {\it symbols} are possible CTTS (see Section\,\ref{Halphasection})
and  {\it diamonds} are 
stars with IR excesses in at least one of the
 four indices.  The {\it dashed lines} indicate the limits
 used to define  the regions with IR excess stars 
 ({\it gray shade regions}).} 
\label{5623fi15}
\end{figure}
\subsection{ H$\alpha$}
For the subsample of  
members for which we also have  
H$\alpha$ spectra, we plotted 
the Q indices 
versus  the FWZI of H$\alpha$, as shown in Fig.\,\ref{5623fi15},
where  
{\it small dots} and {\it large bullets} are WTTS and CTTS, respectively,
defined based on the H$\alpha$ line, while
{\sf x} {\it symbols} are possible CTTS (see Section\,\ref{Halphasection});
{\it diamonds} are stars that show
an IR excess at least in one of the four  indices. 

29 out of the 31 stars classified  as certain CTTS from H$\alpha$ profile show 
IR excesses according to at least one index,
and this is consistent with the presence of an
 accretion disk; 
on the other hand, 
we consider the   
WTTS stars with IR excesses as PMS stars with inert disks.   
The  three stars classified as CTTS according to the H$\alpha$ profile
   that do not show IR excesses
are indicated with the 2MASS flag "$c$", which means that their
photometry is biased by a nearby star that can contaminate the background 
estimation and
therefore their IR magnitudes should be taken with caution. 
No further information can be deduced from Q indices for the  
possible CTTS. 

\subsection{ Binarity}
\begin{figure*}[!ht]
\centerline{\psfig{figure=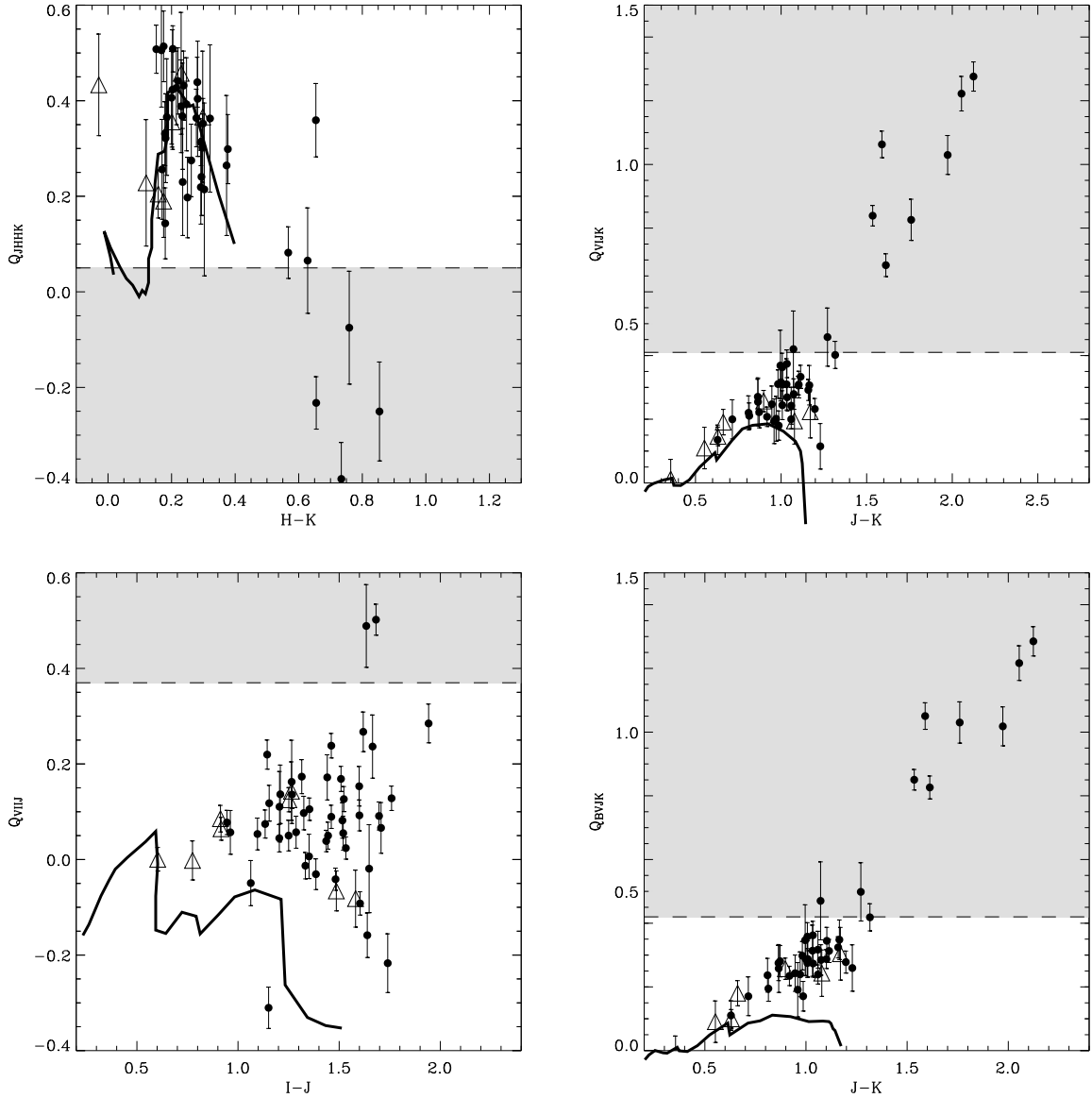}}
\caption{Extinction-free indices of IR excesses for the binaries. 
SB2 stars are indicated with 
{\it dots}, while candidate SB1 stars are indicated with {\it triangles}
(compare with Fig.\,\ref{5623fi12}). The {\it solid line} indicates normal star
colors from \citet{keny95}.  The {\it dashed lines} indicate the limits
 used to define the regions with IR excess stars 
 ({\it gray shade regions}).} 
\label{5623fi16}
\end{figure*}

Fig.\,\ref{5623fi16}  shows the  optical-IR indices for binary stars. 
{\it Dots} are the SB2 stars, while {\it triangles} are the SB1 members (see
Section\,\ref{membership_section}).
Taking into account binaries with  errors on each of the 
IR magnitudes smaller than 0.1 mag, 
we find that  
10\% (4/39) of them show Q$_{\rm JHHK}$ excesses,
22\% (9/41) show Q$_{\rm VIJK}$ and Q$_{\rm BVJK}$ excesses, while only 
6\% (3/47) show Q$_{\rm VIIJ}$ excesses; all the 
 \nsbuno\  candidates SB1 are without IR excesses. The percentage of SB2
 stars with IR excesses in at least one of the four Q indices is 
 25\% (10/40).  
 
 On the other hand,
among the single stars,
21\% (29/137) show Q$_{\rm JHHK}$ excesses,
51\% (74/145) show Q$_{\rm VIJK}$ and Q$_{\rm BVJK}$ excesses, while 
8\% (13/160) show Q$_{\rm VIIJ}$ excesses. The percentage of single stars
 with IR excesses in at least one of the four Q indices is 
 57\% (81/142).
The result supports the hypothesis that companions
do not contribute to the IR excess, as suggested
by \citet{gras05}.  On the contrary, the significantly smaller fraction
of binaries with disks (compared to single stars) indicates that 
binary stars in NGC\,6530 have already undergone a significant disk evolution
and inner disk clearing, probably
due to gravitational effects of the binary components on the
circumstellar disks \citep{armi99}, a theoretical prediction for which no
clear observational evidence has been found until now 
\citep[ and references therein]{moni06}.
\subsection{ Summary}
We summarize in Table\,\ref{5623tab6}, the properties of member
stars, according to the IR excess indices and the accretion
indication from the H$\alpha$ line.
In particular, in the upper Table, 
we give the number of X-ray detected and undetected members,
the number of members with RV at 3\,$\sigma$  consistent and inconsistent 
with cluster membership and the total number of members, for the stars
with IR excesses in the specified index. The numbers corresponding to the
"Q$_{\rm xxxx}$" line indicate the total members having
 IR excess in at least one of the four indices, while 
 the numbers corresponding
 to the line "All" indicate the total numbers 
 of members, regardless of their photometry.
 
Analogously, in the lower Table, we give  the same numbers with respect
to the sample of members with H$\alpha$ observations.

The percentage of members recovered by X-ray detection
is 95\% (174/184), while the members showing IR excesses are 44\% (81/184).
Among X-ray missed members, 7 out of 10 are  recovered using only optical-IR 
photometry; this means that
the percentage of members not recovered by X-ray detections and/or IR excess
diagnostics is equal to about  2\% (3/184).  

If we consider the sample of cluster members with H$\alpha$ observations,
we find that approximately 40\%  of the stars show accreting disks
(profile types PCI, PCIII, YY Ori in Table \ref{5623tab3}). This
is a percentage similar to that found in the Orion Nebula Cluster
by \citet{sici05} with similar indicators.
\begin{table}
\centering
\tabcolsep 0.2truecm
\caption{{\it Top:} Number of X-ray detected and undetected members,
with RV at 3\,$\sigma$  consistent and unconsistent 
with cluster membership and total number of members, for the stars
with IR excesses in the specified index. The numbers corresponding to the
"Q$_{xxxx}$" line, indicate the total members having
 IR excess in at least one of the four indices, while,
 the numbers corresponding
 to the line "All" indicate the total of members, regardless their photometry.
{\it Bottom:} same numbers with respect
to the sample of members with H$_\alpha$ observations.}
\vspace{0.8cm}
\begin{tabular}{l*{1}{c}}
\makebox[18em]{Total single member sample}\\
\end{tabular}
\begin{tabular}{|l||*{6}{c|}}\hline
\makebox[3em]{}
&\makebox[3em]{X-ray}&\makebox[3em]{No X-ray}
&\makebox[3em]{RV}&\makebox[3em]{No RV}&\makebox[3em]{Tot.}\\\hline\hline
Q$_{\rm JHHK}$&          25&           4&          29&           0&
          29\\\hline
Q$_{\rm VIJK}$&          67&           7&          74&           0&
          74\\\hline
Q$_{\rm VIIJ}$&          11&           2&          13&           0&
          13\\\hline
Q$_{\rm BVJK}$&          68&           6&          74&           0&
          74\\\hline
Q$_{\rm xxxx}$&          74&           7&          81&       0&          81
\\\hline
All     &         174&          10&         176&           8&         184
\\\hline
\end{tabular}

\vspace{0.5cm}
\begin{tabular}{l*{1}{c}}
\makebox[18em]{Single member sample with $H_\alpha$ obs.}\\
\end{tabular}
\begin{tabular}{|l||*{6}{c|}}\hline
\makebox[3em]{}
&\makebox[3em]{X-ray}&\makebox[3em]{No X-ray}
&\makebox[3em]{RV}&\makebox[3em]{No RV}&\makebox[3em]{Tot.}\\\hline\hline
CTTS &          29&           2&          31&           0&          31\\\hline
All  &          67&           4&          69&           2&          71\\\hline
\end{tabular}
\label{5623tab6}
\end{table}

\section{Lithium abundances and depleted members \label{alisection}}
As mentioned in Section\,\ref{membership_section}, we consider as cluster  
members all  objects with lithium EW larger than the adopted threshold
(see {\it dashed line} in Fig.\,\ref{5623fig9}), having a RV consistent with 
the cluster membership and/or  an X-ray counterpart. 
The adopted membership
criteria always require the presence of the lithium line, because this
condition has to be always true for young stars, such as those of  
NGC\,6530.  
In the sample of certain cluster members we also included X-ray 
detected
stars, with RV consistent with the cluster membership and EW of the lithium
measurable but
smaller than the EW threshold.

While the lithium EWs strongly depend on temperature, 
the lithium abundance is 
expected to be constant  for very young
stars such as our targets, which should not have yet reached interior
temperatures for  lithium burning.   

However, accurate lithium abundances cannot be derived from our data due to the
large uncertainty affecting both our temperature estimates and lithium EW
measurements.
The large spread in temperature is due to the large errors mainly introduced by 
unknown individual reddening values, while the spread in the EWs is mainly 
due to the continuum normalization and to the veiling, which has
been estimated to be of the same order as the random errors affecting the EW measures.

In addition, the growth curves given in the literature 
to derive lithium abundances from EWs 
are available only for stars with $4000<$T$<6500$\,K \citep{sode93}, while
our sample contains stars with
 temperatures down to T$\sim$3500\,K, where the physics
of stellar atmospheres is not well known due to the presence of
molecular lines such as TiO and H$_2$O. A further uncertainty is introduced
by non--local thermodynamic equilibrium (NLTE)
\citep{carl94}.

Using our data, the \citet{sode93} growth curves and the NLTE lithium correction
of \citet{carl94},  we find that, for the sample of stars classified as members
and $T>4000$\,K, the median value of the lithium abundance is A(Li)=3.0
dex\footnote{A(Li)=log($n_{Li}/n_H$)+12 \citep{ande89}},
with a standard deviation of 0.6 dex. 
While the median value is consistent with the expected cosmic lithium abundance,
the spread is quite large and is mainly due to the few objects
that have a lithium EW smaller than those of similar temperature stars.
Based on the EW values of the bulk of our data, and taking the 
EW of the Pleiades
as a reference for older stars, we consider the stars with EWs larger than the
{\it dashed line} in Fig.\,\ref{5623fig9} 
and temperature smaller than about 5000\,K
as lithium-undepleted cluster members.
We also consider as lithium-undepleted members
the few objects with temperature higher than 5000\,K.

\begin{figure}[!ht]
\centerline{\psfig{figure=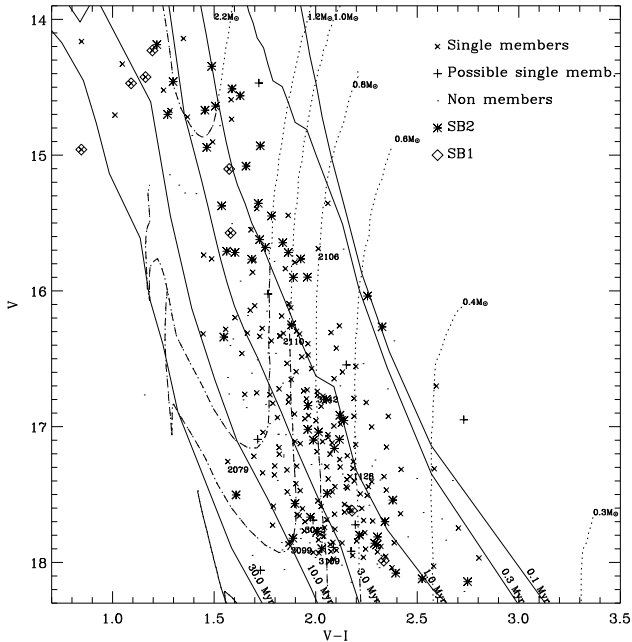,width=9cm}}
\caption{V vs. V-I color-magnitude diagram of candidate members of
NGC\,6530. Theoretical isochrones and
tracks of \citet{sies00} are drawn and different levels of predicted Li depletion are
also indicated; portions of the tracks 
with undepleted Li are drawn as {\it dotted lines}, those with predicted Li depletion
up to a factor of 10 are drawn as {\it dotted-dashed lines} and those with
predicted Li depletion more than a factor of 10 are drawn as {\it solid lines}.
The 8 stars with evidence
 of Li depletion and/or veiling effects are also indicated.}
\label{5623fi17}
\end{figure}

Particular attention has been devoted to  the X-ray detected
objects, with a RV consistent with cluster membership and EW
smaller than the limit indicated by the {\it dashed line}, as 
shown in Fig.\,\ref{5623fig9}.
Table\,\ref{5623tab7} lists the values obtained in this paper for 
these stars; the estimated veiling corrections (r) are also reported
for the stars in which the three lines adopted to estimate 
 the veiling  are visible. Using these values to correct our measured
 EW of the lithium line, we find that the corrected EWs are consistent with
the values of the Li EW  of similar temperature members for the stars
with spectrum numbers Sp 1128, 3042 and 3099, while, for the star  
 Sp 2079, it remains smaller  than the adopted threshold.
 Note that the r values are among the highest values in Fig.\,\ref{5623fi13}.
 The age estimated for the star Sp 2079 (15\,Myr) is consistent with a significant 
lithium depletion, as predicted by theoretical models (see 
Fig.\,\ref{5623fi17}).
However, this value can be affected
by a systematic error due to the veiling, which can modify the photospheric
colors of the stars. In addition, its temperature, among the
highest in our sample, is affected by a large error
due to the unknown reddening.
 We do not exclude the hypothesis  
that this object is a young main sequence field star.

 For the remaining 4 stars, 
 only the lithium line is clearly visible and this suggest that they
 are highly veiled objects. 
 This hypothesis can be confirmed for the
 stars 2106  and 2110, for which evidence of accretion has been found from the
H$\alpha$ line.  The presence of circumstellar disks, deduced by IR excesses,
can be considered as a further indication of possible accretion phenomena
which can cause significant veiling.

The position of all our sample stars 
in the  observed $V$ vs. $V-I$ color-magnitude diagram 
 is shown in Fig.\,\ref{5623fi17},  
where the dereddened theoretical isochrones and tracks of \citet{sies00} are drawn.
 The position
of the 8 stars with low Li EWs  and the 
regions with expected lithium depletion are also indicated;
portions of the tracks 
with undepleted Li are drawn as {\it dotted lines}, those with predicted Li depletion
up to a factor of 10 are drawn as {\it dotted-dashed lines} and those with
predicted Li depletion more than a factor of 10 are drawn as {\it solid lines.}

Temperature estimates from spectral types,
 allowing a consistent comparison with theoretical models,
will be helpful to establish whether these stars are actually lithium depleted
or if the EW of these stars are smaller than those of similar temperature 
stars because of only veiling  effects, or whether they are young field stars.

\begin{table*}
\centering
\tabcolsep 0.1truecm
\caption{Properties of stars with evidence of Li depletion and/or veiling effects.}
\vspace{0.5cm}
\begin{tabular}{cccccccccccccccccccc}
\hline
\hline
\\
\multicolumn{1}{c}{Sp}&
\multicolumn{1}{c}{Age}&
\multicolumn{1}{c}{M}&
\multicolumn{1}{c}{vsin(i)}&
\multicolumn{1}{c}{$\sigma_{\rm vsin(i)}$}&
\multicolumn{1}{c}{T$_{\rm eff}$}&
\multicolumn{1}{c}{Li\,{\scriptsize I}}&
\multicolumn{1}{c}{r\tablenotemark{\it a}}&
\multicolumn{1}{c}{H$_\alpha$}&
\multicolumn{1}{c}{X-ray}&
\multicolumn{1}{c}{RV}&
\multicolumn{1}{c}{Q}&
\multicolumn{1}{c}{Q}&
\multicolumn{1}{c}{Q}&
\multicolumn{1}{c}{Q}&
\multicolumn{1}{c}{H$_\alpha$}&
\multicolumn{1}{c}{H$_\alpha$}&
\multicolumn{1}{c}{Type}\\
&(Myr)&($M_\odot$)&[km/s]&[km/s]&K&{\tiny EW[m$\AA$]}&&{\tiny FWZI[$\AA$]}&&&
{\tiny \it JHHK}&{\tiny \it VIJK}&{\tiny \it VIIJ}&{\tiny \it BVJK}&CTTS&PT& \\ 
\hline\\ 
 1128&  1.06 &  0.61 &        21.9 &         2.0 &    3902&     420.& 0.3& -- &     Y&     Y&     Y&     Y&     Y&     Y&    --&    --&     M\\
 2079& 15.03 &  1.11 &        10.8 &         0.5 &    4648&     176.& 0.5& 2.2&     Y&     Y&    --&    --&     N&    --&  WTTS&    AE&     M\\
 2106&  0.44 &  0.81 &        30.2 &         4.6 &    4108&     268.& 1.9&15.6&     Y&     Y&     Y&     Y&     N&     Y&  CTTS&   PCI&     M\\
 2110&  1.35 &  1.06 &        59.6 &         7.7 &    4319&     307.& NaN&13.2&     Y&     Y&     Y&     Y&     Y&     Y&  CTTS& PCIII&     M\\
 3013&  5.42 &  0.94 &  $\le 10.0$ &             &    4190&     220.& NaN& -- &     Y&     Y&     N&     Y&     Y&     Y&    --&    --&     M\\
 3042&  1.12 &  0.79 &  $\le 10.0$ &             &    4063&     366.& 0.3& -- &     Y&     Y&     Y&     Y&     Y&     Y&    --&    --&     M\\
 3099&  9.48 &  1.00 &        35.2 &         5.1 &    4272&     345.& 0.7& -- &     Y&     Y&    --&    --&     N&    --&    --&    --&     M\\
 3109&  5.47 &  0.85 &  $\le 10.0$ &             &    4104&     322.& 1.9& -- &     Y&     Y&     N&     Y&     N&     Y&    --&    --&     M\\
\hline
&&&&&&&&&&&&&&&&&&&\\
\multicolumn{4}{c}{$^a$ Typical error on the veiling correction is 0.3}&&&&&&&&&&&&&&\\
\end{tabular}
\label{5623tab7}
\end{table*}

\section{Summary and conclusions}
We used  spectra of intermediate resolution (R$\sim$19\,300), taken with 
the GIRAFFE-FLAMES multi-object spectrograph of the ESO-VLT Kueyen Telescope,
to analyze a
sample of \nallobserved\ candidate members of the
young cluster NGC\,6530. The targets were selected in the region of  PMS
stars of the $V$ vs. $V-I$ color-magnitude diagram, traced
by a large number of X-ray detected stars \citep{pris05}.
72\% of the selected targets are X-ray active stars.

The spectra of these stars allowed us to measure both radial and rotational
velocities using the technique of Cross-Correlation. 
This technique enabled us also to identify 53 objects
as binaries. In addition we measured the EW of the lithium
line, used as a membership criterion. H$\alpha$
spectra, available  for a subsample of 115 stars, allowed us also
to select WTTS and CTTS.

The measures obtained from our spectra were used to 
 select a sample of certain and possible members,
based on the combination of the RV measures, the Li EWs
and the X-ray detection. We find \nlicert\ certain single members
that, added to the \nbinaries\ binaries, gives us a total sample
of \nmembers\ certain members plus 10 possible members.

We used extinction-free indices to select stars with IR excesses,
indicators of the presence of circumstellar disks. 
A statistically significant difference was found between the 
rotational velocity distributions
of K and M type stars  with IR excesses with respect to that of
the same spectral type stars but
 without IR excesses. In addition we find that
all stars, except three, classified as CTTS based on the H$\alpha$ line
have IR 
excesses and this confirms that most of the IR excesses
stars have accretion phenomena.  These results are consistent with 
the disk locking picture suggesting a rotational velocity reduction
in stars with disks, as already found in the Orion Nebula Cluster
\citep[][ and reference therein]{sici05,rebu06}.

We also find that the fraction of binaries with disks is significantly
smaller than that of single IR excess stars and this suggest that gravitational
effects of the 
binary components in  NGC\,6530 caused a  significant disk evolution
and disk clearing.

A fraction equal to about 44\% (81/184)
of the stars identified as cluster members show IR excesses and, among the
X-ray undetected members,
about 70\% (7/10) have IR excess,
recovered using only optical-IR photometry.
This confirms the importance of using both X-ray data and 
optical-IR data as diagnostics of  membership.

The EWs of the lithium line of the stars classified as cluster members are
significantly larger than those of older stars (Pleiades like) of similar
temperature and this is agreement with the young ages of NGC\,6530. We computed
the lithium abundances of cluster members; while the median value of these
values, equal to A(Li)=3.0 dex, is in agreement with the predicted cosmic
abundance, the rms, equal to 0.6 dex, is quite large; this is due to the 
uncertainty in the computed abundance values due to the  uncertainties affecting
the temperature estimates, the lithium EW  measures (including the veiling
correction), the adopted growth curves and the non-local thermodynamic
equilibrium.

We find 8 possible cluster members that are X-ray detected stars
and with RV consistent with cluster membership,  with Li EW
smaller than the adopted threshold.  For 7 of them, we find evidence
of strong veiling, which can explain their small Li EW values, while
for one we find that the veiling corrected EW is however smaller than
those of stars of similar temperature. The age estimated for this star
is consistent with a significant lithium depletion, but we do not 
exclude  that possible systematic errors due to the veiling can
affect the determination of stellar parameters (including age). 

\begin{acknowledgements}
We wish to thank Piercarlo Bonifacio for his contribution
 in taking FLAMES observations
and Paolo Span\`o for his help in some steps of data analysis.
We also wish to thank the anonymous referee for useful comments
and suggestions.
We acknowledge financial support from the Italian MIUR.  
\end{acknowledgements}
\bibliographystyle{aa}
\bibliography{5623}

\begin{thebibliography}{38}
\expandafter\ifx\csname natexlab\endcsname\relax\def\natexlab#1{#1}\fi

\bibitem[{{Anders} \& {Grevesse}(1989)}]{ande89}
{Anders}, E. \& {Grevesse}, N. 1989, \gca, 53, 197

\bibitem[{{Arias} {et~al.}(2006){Arias}, {Barb{\'a}}, {Ma{\'{\i}}z
  Apell{\'a}niz}, {Morrell}, \& {Rubio}}]{aria06}
{Arias}, J.~I., {Barb{\'a}}, R.~H., {Ma{\'{\i}}z Apell{\'a}niz}, J., {Morrell},
  N.~I., \& {Rubio}, M. 2006, \mnras, 366, 739

\bibitem[{{Armitage} {et~al.}(1999){Armitage}, {Clarke}, \& {Tout}}]{armi99}
{Armitage}, P.~J., {Clarke}, C.~J., \& {Tout}, C.~A. 1999, \mnras, 304, 425

\bibitem[{{Bertout}(1984)}]{bert84}
{Bertout}, C. 1984, Reports of Progress in Physics, 47, 111

\bibitem[{{Carlsson} {et~al.}(1994){Carlsson}, {Rutten}, {Bruls}, \&
  {Shchukina}}]{carl94}
{Carlsson}, M., {Rutten}, R.~J., {Bruls}, J.~H.~M.~J., \& {Shchukina}, N.~G.
  1994, \aap, 288, 860

\bibitem[{{Cutri} {et~al.}(2003)}]{cutr03}
{Cutri}, R.~M. {et~al.} 2003, "2MASS All-Sky Catalog of Point Sources",
  University of Massachusetts and Infrared Processing and Analysis Center,
  (IPAC/CAlifornia Institute of Technology.

\bibitem[{{Damiani} {et~al.}(2004){Damiani}, {Flaccomio}, {Micela},
  {Sciortino}, {Harnden}, \& {Murray}}]{dami04}
{Damiani}, F., {Flaccomio}, E., {Micela}, G., {et~al.} 2004, \apj, 608, 781

\bibitem[{{Damiani} {et~al.}(2006){Damiani}, {Prisinzano}, {Micela}, \&
  {Sciortino}}]{dami06}
{Damiani}, F., {Prisinzano}, L., {Micela}, G., \& {Sciortino}, S. 2006, \aap,
  in press

\bibitem[{{D'Antona} \& {Mazzitelli}(1997)}]{dant97}
{D'Antona}, F. \& {Mazzitelli}, I. 1997, Memorie della Societa Astronomica
  Italiana, 68, 807

\bibitem[{{Dullemond} {et~al.}(2006){Dullemond}, {Natta}, \& {Testi}}]{dull06}
{Dullemond}, C.~P., {Natta}, A., \& {Testi}, L. 2006, \apjl, 645, L69

\bibitem[{{Edwards} {et~al.}(1993){Edwards}, {Strom}, {Hartigan}, {Strom},
  {Hillenbrand}, {Herbst}, {Attridge}, {Merrill}, {Probst}, \&
  {Gatley}}]{edwa93}
{Edwards}, S., {Strom}, S.~E., {Hartigan}, P., {et~al.} 1993, \aj, 106, 372

\bibitem[{{Gras-Vel{\'a}zquez} \& {Ray}(2005)}]{gras05}
{Gras-Vel{\'a}zquez}, {\`A}. \& {Ray}, T.~P. 2005, \aap, 443, 541

\bibitem[{{Herbst} {et~al.}(2002){Herbst}, {Bailer-Jones}, {Mundt},
  {Meisenheimer}, \& {Wackermann}}]{herb02}
{Herbst}, W., {Bailer-Jones}, C.~A.~L., {Mundt}, R., {Meisenheimer}, K., \&
  {Wackermann}, R. 2002, \aap, 396, 513

\bibitem[{{Herbst} \& {Mundt}(2005)}]{herb05}
{Herbst}, W. \& {Mundt}, R. 2005, \apj, 633, 967

\bibitem[{{Herbst} {et~al.}(2000){Herbst}, {Rhode}, {Hillenbrand}, \&
  {Curran}}]{herb00}
{Herbst}, W., {Rhode}, K.~L., {Hillenbrand}, L.~A., \& {Curran}, G. 2000, \aj,
  119, 261

\bibitem[{{Jeffries} \& {Oliveira}(2005)}]{jeff05}
{Jeffries}, R.~D. \& {Oliveira}, J.~M. 2005, \mnras, 358, 13

\bibitem[{{Kenyon} \& {Hartmann}(1995)}]{keny95}
{Kenyon}, S.~J. \& {Hartmann}, L. 1995, \apjs, 101, 117

\bibitem[{{Koenigl}(1991)}]{koen91}
{Koenigl}, A. 1991, \apjl, 370, L39

\bibitem[{{Kurucz}(1993)}]{kuru93}
{Kurucz}, R.~L. 1993, {SYNTHE spectrum synthesis programs and line data}
  (Kurucz CD-ROM, Cambridge, MA: Smithsonian Astrophysical Observatory, |c1993,
  December 4, 1993)

\bibitem[{{Lada} {et~al.}(1976){Lada}, {Gottlieb}, {Gottlieb},
  {et~al.}}]{lada76}
{Lada}, C.~J., {Gottlieb}, C.~A., {Gottlieb}, E.~W., {et~al.} 1976, \apj, 203,
  159

\bibitem[{{Monin} {et~al.}(2006){Monin}, {Clarke}, {Prato}, \&
  {McCabe}}]{moni06}
{Monin}, J.~L., {Clarke}, C.~J., {Prato}, L., \& {McCabe}, C. 2006, Protostars
  and Planets V, in press

\bibitem[{{Munari} \& {Carraro}(1996)}]{muna96}
{Munari}, U. \& {Carraro}, G. 1996, \aap, 314, 108

\bibitem[{{Palla} {et~al.}(2005){Palla}, {Randich}, {Flaccomio}, \&
  {Pallavicini}}]{pall05}
{Palla}, F., {Randich}, S., {Flaccomio}, E., \& {Pallavicini}, R. 2005, \apjl,
  626, L49

\bibitem[{{Prisinzano} {et~al.}(2005){Prisinzano}, {Damiani}, {Micela}, \&
  {Sciortino}}]{pris05}
{Prisinzano}, L., {Damiani}, F., {Micela}, G., \& {Sciortino}, S. 2005, \aap,
  430, 941

\bibitem[{{Randich} {et~al.}(2001){Randich}, {Pallavicini}, {Meola},
  {Stauffer}, \& {Balachandran}}]{rand01}
{Randich}, S., {Pallavicini}, R., {Meola}, G., {Stauffer}, J.~R., \&
  {Balachandran}, S.~C. 2001, \aap, 372, 862

\bibitem[{{Rauw} {et~al.}(2002){Rauw}, {Naz{\' e}}, {Gosset}, {Stevens},
  {Blomme}, {Corcoran}, {Pittard}, \& {Runacres}}]{rauw02}
{Rauw}, G., {Naz{\' e}}, Y., {Gosset}, E., {et~al.} 2002, \aap, 395, 499

\bibitem[{{Rebull} {et~al.}(2006){Rebull}, {Stauffer}, {Megeath}, {Hora}, \&
  {Hartmann}}]{rebu06}
{Rebull}, L.~M., {Stauffer}, J.~R., {Megeath}, S.~T., {Hora}, J.~L., \&
  {Hartmann}, L. 2006, \apj, 646, 297

\bibitem[{{Rhode} {et~al.}(2001){Rhode}, {Herbst}, \& {Mathieu}}]{rhod01}
{Rhode}, K.~L., {Herbst}, W., \& {Mathieu}, R.~D. 2001, \aj, 122, 3258

\bibitem[{{Rieke} \& {Lebofsky}(1985)}]{riek85}
{Rieke}, G.~H. \& {Lebofsky}, M.~J. 1985, \apj, 288, 618

\bibitem[{{Sicilia-Aguilar} {et~al.}(2005){Sicilia-Aguilar}, {Hartmann},
  {Szentgyorgyi}, {Fabricant}, {F{\H u}r{\'e}sz}, {Roll}, {Conroy}, {Calvet},
  {Tokarz}, \& {Hern{\'a}ndez}}]{sici05}
{Sicilia-Aguilar}, A., {Hartmann}, L.~W., {Szentgyorgyi}, A.~H., {et~al.} 2005,
  \aj, 129, 363

\bibitem[{{Siess} {et~al.}(2000){Siess}, {Dufour}, \& {Forestini}}]{sies00}
{Siess}, L., {Dufour}, E., \& {Forestini}, M. 2000, \aap, 358, 593

\bibitem[{{Soderblom} {et~al.}(1993){Soderblom}, {Jones}, {Balachandran},
  {Stauffer}, {Duncan}, {Fedele}, \& {Hudon}}]{sode93}
{Soderblom}, D.~R., {Jones}, B.~F., {Balachandran}, S., {et~al.} 1993, \aj,
  106, 1059

\bibitem[{{Stassun} {et~al.}(1999){Stassun}, {Mathieu}, {Mazeh}, \&
  {Vrba}}]{stas99}
{Stassun}, K.~G., {Mathieu}, R.~D., {Mazeh}, T., \& {Vrba}, F.~J. 1999, \aj,
  117, 2941

\bibitem[{{Sung} {et~al.}(2000){Sung}, {Chun}, \& {Bessell}}]{sung00}
{Sung}, H., {Chun}, M., \& {Bessell}, M.~S. 2000, \aj, 120, 333

\bibitem[{{The}(1960)}]{the60}
{The}, P. 1960, \apj, 132, 40

\bibitem[{{Tonry} \& {Davis}(1979)}]{tonr79}
{Tonry}, J. \& {Davis}, M. 1979, \aj, 84, 1511

\bibitem[{{van den Ancker} {et~al.}(1997){van den Ancker}, {The}, {Feinstein},
  {Vazquez}, {de Winter}, \& {Perez}}]{anck97}
{van den Ancker}, M.~E., {The}, P.~S., {Feinstein}, A., {et~al.} 1997, \aaps,
  123, 63

\bibitem[{{Walker}(1957)}]{walk57}
{Walker}, M.~F. 1957, \apj, 125, 636

\end{thebibliography}
\end{document}